\documentclass[aps,pra,showpacs,amssymb,superscriptaddress,nofootinbib,twocolumn]{revtex4-1}
\usepackage[ansinew]{inputenc}
\usepackage{bbm}
\usepackage{bm}
\usepackage{amsbsy}
\usepackage{amsthm}
\usepackage{amssymb}
\usepackage{amsfonts}
\usepackage{amsmath}
\usepackage{dsfont} 
\usepackage{graphicx} 
\usepackage{epsfig}
\usepackage{epstopdf}
\usepackage{dsfont}
\usepackage{color}
\usepackage[colorlinks]{hyperref}
\usepackage[figure,table]{hypcap}
\usepackage{enumerate}
\hypersetup{
	bookmarksnumbered,
	pdfstartview={FitH},
	citecolor={darkgreen},
	linkcolor={darkred},
	urlcolor={darkblue},
	pdfpagemode={UseOutlines}}
\definecolor{darkgreen}{RGB}{50,190,50}
\definecolor{darkblue}{RGB}{0,0,190}
\definecolor{darkred}{RGB}{238,0,0}
\usepackage{soul}
\newcommand{\pr}{^{\prime}}
\newcommand{\ket}[1]{\ensuremath{\left|{#1}\right\rangle}}

\newcommand{\bra}[1]{\ensuremath{\left\langle{#1}\right|}}

\newcommand{\piP}[1]{\ensuremath{\pi_{\protect\raisebox{0pt}{\hspace*{-1pt}\tiny{$#1$}}}}}
\newcommand{\pitildeP}[1]{\ensuremath{\tilde{\pi}_{\protect\raisebox{0pt}{\hspace*{-1pt}\tiny{$#1$}}}}}

\DeclareMathOperator{\diag}{diag}

\newcommand{\djj}{d\kern-0.4em\char"16\kern-0.1em}

\begin{document}

\title{Quantum-enhanced deliberation of learning agents using trapped ions}
\author{Vedran Dunjko}
\email{vedran.dunjko@uibk.ac.at}
\affiliation{
Institute for Quantum Optics and Quantum Information,
Austrian Academy of Sciences,
Technikerstra{\ss}e 21a,
A-6020 Innsbruck,
Austria}
\affiliation{
Institute for Theoretical Physics, University of Innsbruck,
Technikerstra{\ss}e 25,
A-6020 Innsbruck,
Austria}
\affiliation{Laboratory of Evolutionary Genetics, Division of Molecular Biology,
Ru{\djj}er Bo\v{s}kovi{\'c} Institute, Bijeni\v{c}ka cesta 54, HR-10000 Zagreb, Croatia}
\author{Nicolai Friis}
\email{nicolai.friis@uibk.ac.at}
\affiliation{
Institute for Quantum Optics and Quantum Information,
Austrian Academy of Sciences,
Technikerstra{\ss}e 21a,
A-6020 Innsbruck,
Austria}
\author{Hans J.~Briegel}
\email{hans.briegel@uibk.ac.at}
\affiliation{
Institute for Quantum Optics and Quantum Information,
Austrian Academy of Sciences,
Technikerstra{\ss}e 21a,
A-6020 Innsbruck,
Austria}
\affiliation{
Institute for Theoretical Physics, University of Innsbruck,
Technikerstra{\ss}e 25,
A-6020 Innsbruck,
Austria}
\date{\today}
\begin{abstract}
A~scheme that successfully employs quantum mechanics in the design of autonomous learning agents has recently been reported in the context of the projective simulation (PS) model for artificial intelligence. In that approach, the key feature of~a PS agent,~a specific type of memory which is explored via random walks, was shown to be amenable to quantization, allowing for~a speed-up. In this work we propose an implementation of such classical and quantum agents in systems of trapped ions. We employ~a generic construction by which the classical agents are `upgraded' to their quantum counterparts by~a nested process of adding coherent control, and we outline how this construction can be realized in ion traps. Our results provide~a flexible modular architecture for the design of PS agents. Furthermore, we present numerical simulations of simple PS agents which analyze the robustness of our proposal under certain noise models.
\end{abstract}

\maketitle
\section{Introduction}\label{sec:intro}
In the past decades, quantum physics has been employed to enhance communication and information processing with significant success, laying the foundation for the now well established fields of quantum computation and quantum information~\cite{Deutsch1985,DeutschJozsa1992,Grover1996,Shor1994,NielsenChuang2000}. In contrast, the potential of merging the related, but distinct, field of artificial intelligence (AI) with quantum physics is significantly less well-understood. Thus far, advances in this field have been reported mostly for algorithmic approaches to applied AI-related tasks, e.g., (un-)supervised data clustering and process replication, where selected quantum algorithms could be utilized~\cite{NevenDenchevRoseMacready2008, LloydMohseniRebentrost2013,ManzanoPawlowskiBrukner2009,PudenzLidar2013,AimeurBrassardGambs2013}.

On the other hand, the first result showing that quantum mechanics can also aid in the complemental task of designing autonomous learning agents\textemdash a~task more closely related to robotics, and embodied cognitive sciences\textemdash has only recently been provided in Ref.~\cite{PaparoDunjkoMakmalMartinDelgadoBriegel2014}. This work is embedded in the framework of projective simulation (PS) for AI, introduced in Ref.~\cite{BriegelDeLasCuevas2012}. The central component of PS is~a specific memory system utilized by the agent. This memory system, called episodic and compositional memory (ECM), provides~a platform for \emph{simulating future action} before real action is taken. The ECM can be described as a stochastic network of so-called \emph{clips}, which represent prior experiences of the learning agent, whose decision-making process is realized by~a stochastic random walk in the clip space. In the agent's design, it is the specific structure of the ECM that is particularly suitable for quantization.

In this work we present~a proposal for the experimental implementation of both classical and quantum PS agents in systems of trapped ions. While the classical variants of PS agents can easily be realized in physical systems without requiring quantum control, we show here how certain implementations of classical agents in ion traps can be used to construct quantum PS agents. This is achieved in~a generic way through~a nested process of adding \emph{coherent control}.

The outline of this paper is as follows. In Section~\ref{sec:PS} we briefly review the PS model and give the basic operational elements which have to be constructed in an implementation of~a classical or quantum PS agent. Then, in Section~\ref{sec:SPS} we give~a more formal treatment of the standard, classical PS agent, and show explicitly how such an agent may be implemented in an ion trap set-up. In particular, in Section~\ref{subsec:Coherent Controlization in Trapped Ions}, we discuss how the technique of adding coherent control provides~a generic construction for emulating the standard PS agent in quantum systems, specifically in trapped ions. Finally, in Section~\ref{sec:reflecting projective simulators in ion traps}, we extend our analysis to quantum PS agents by specifying all required operations and describing their implementation in ion traps. In the \hyperref[Appendix]{Appendix} we further present~a simple example for~a quantum PS agent that can be straightforwardly implemented in an ion trap, for which we provide numerical simulations incorporating an appropriate error model.\\
\vspace*{-6mm}

\section{Projective Simulation}\label{sec:PS}
\begin{figure}[ht!]
\includegraphics[width=0.42\textwidth]{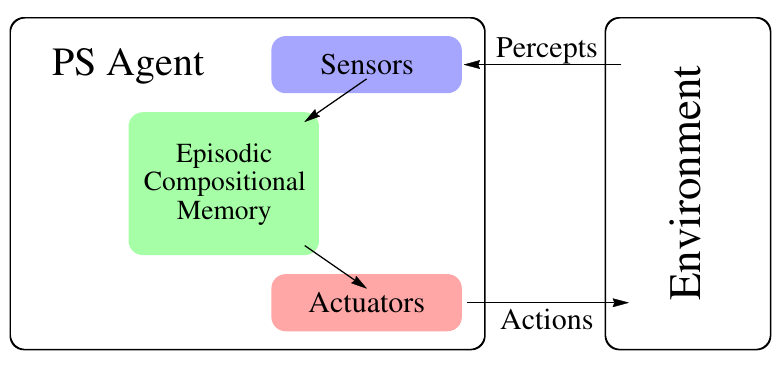}
\caption{\label{fig:learning agent schematic}
\textbf{Projective simulation agent}. The (PS) model for active learning agents, introduced in Ref.~\cite{BriegelDeLasCuevas2012}, describes an embodied agent that interacts with its environment via sensory input (percepts), and action on the environment that is conducted using~a set of actuators. The sensors and actuators are linked to the episodic compositional memory (ECM), which relates new perceptual input to the agent's past experience.}
\end{figure}

The central component of~a PS agent, illustrated in Fig.~\ref{fig:learning agent schematic}, is the episodic and compositional memory, which can be formally represented as~a stochastic network of \emph{clips}. Clips represent the units of episodic memory, which consist of memorized percepts, actions and ensuing rewards. The process of projective simulation is triggered
\vspace*{-6mm}

\noindent
by perceptual input that initiates~a \emph{random walk} over the clip space. This walk constitutes the stochastic replay of previously established memories and precedes the initiation of real action. The agent's capability to learn is represented by two mechanisms, (i)~the adaption of the transition probabilities between the clips, and (ii)~the addition of new clips under compositional principles.

\begin{figure}[ht!]
\includegraphics[width=0.31\textwidth]{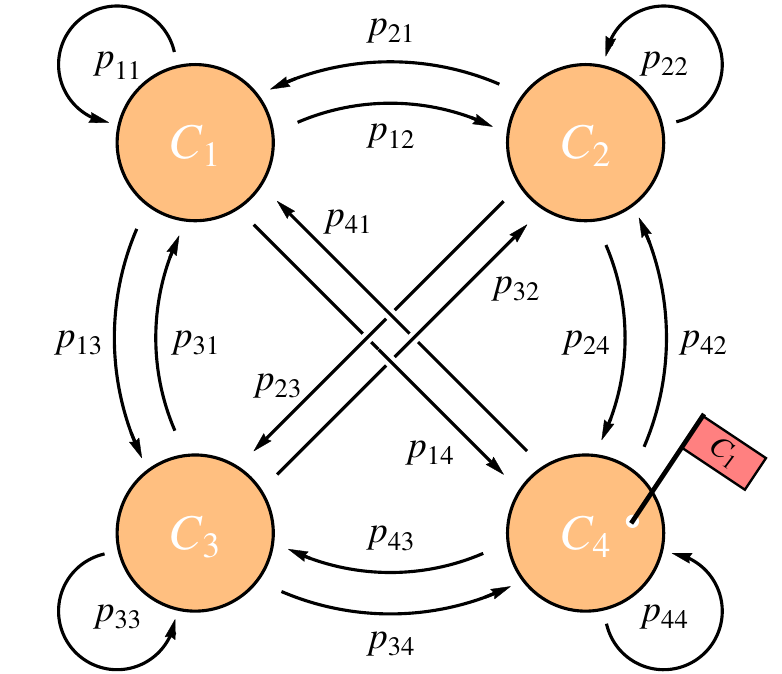}
\caption{\label{fig:clip network}
\textbf{Clip network}. An example for~a network with four clips~$c_{i}\ (i=1,2,3,4)$ is shown. At~any fixed time, the PS agent associates~a discrete-time homogeneous Markov chain with transition matrix~$P=[p_{ij}]\ (i,j=1,2,3,4)$ to the ECM, which governs the transition probabilities for~a random walk in the network. In addition, flags, here indicated on clip $c_{4}$, may be introduced, e.g., to relate actions that were recently rewarded to the corresponding percepts.}
\end{figure}

More formally, at any instance of time the ECM of an agent can be represented as~a directed weighted graph, where the vertices represent the clips, and the weights of the edges represent the transition probabilities, see Fig.~\ref{fig:clip network}. We refer to this graph as the clip network. The random walk, or equivalently, the Markov chain, associated to the process of projective simulation is carried out over the clip network. Finally, the learning aspect of the agent is realized by updating the clip network based on the (re)actions and rewards of the environment, with which it interacts.

The criteria under which an action, that is,~a clip representing~a single memorized action in the ECM, is coupled out as real action can vary, leading to distinct types of PS agents. Here we list a few examples that we will encounter again later in this paper. In the so-called \emph{standard} PS model, the first action clip that is encountered during the random walk over the clip network is coupled out as the chosen real action. The standard PS model can further be equipped with \emph{emotion clips}, which are clip tags indicating, for instance,
whether~a chosen action recently lead to~a reward. In this extended model, the random walk process can be iterated if the encountered action clip carries~a `negative' association \textemdash a~process we will refer to as \emph{reflection}.

Elaborating on the notion of reflection, in Ref.~\cite{PaparoDunjkoMakmalMartinDelgadoBriegel2014} some of the authors have recently introduced the \emph{reflecting PS} (RPS) agent model, in which the Markov chain associated to the clip network is \emph{ergodic}, and hence has a unique stationary distribution over the clip network. The RPS agent continues the random walk until the stationary distribution is reached, and (iteratively) samples from it until an action clip is observed. Building on the approaches of Refs.~\cite{Szegedy2004,MagniezNayakRolandSantha2011} for quantizing random walks, this particular model was shown to have~a quantum analog, called \emph{quantum RPS}, which permits~a quadratic speed-up in active learning scenarios.

As we have mentioned previously, the PS model can be endowed with additional structures, such as the aforementioned emotion tags, which further improve the agent's learning capacity, see Ref.~\cite{MautnerMakmalManzanoTierschBriegel2014}. These additional structures are, in principle, compatible with the constructions we present, but we shall only utilize the simplest of these extensions, so-called flags, in the examples that are considered in the \hyperref[Appendix]{Appendix}. As we will discuss, these flags allow for the demonstration of~a quantum speed-up when incorporated into~a very simple agent design, which is readily implementable in current laboratories.

In the next section, we present a more formal treatment of the standard PS model, and show how it can be implemented in an ionic set-up.
\section{Standard PS agent}\label{sec:SPS}
As noted, in the PS model, the ECM is represented as~a clip network, that is,~a weighted directed graph over the set of vertices $\mathcal{C}=\{c_{i}\}_{i=1}^{N}$, where each~$c_i$ represents~a clip. The directed weighted edges of the graph represent the transition probabilities from one clip to another~\footnote{Technically, since in the standard PS model, the action is coupled out whenever an action clip is hit, the probabilities of transiting from an action clip are undefined. However, we can, for simplicity, assign a unit probability of transiting to itself to each action clip. Thus, action clips are the absorbing states of the underlying Markov chain, although this will not be relevant for our work.} given by~a transition matrix $P=[p_{ij}]$ which is an $N\times N$

\noindent
left-stochastic matrix, that is,~$0\leq p_{ij}\leq1$ and~$\sum_{i}p_{ij}=1\,\forall\,j\,$. In the standard PS, we can assume the clip network always contains clips which are representations of individual percepts (from the set of percepts $\mathcal{S} = \{s_{i}\}_i$) as well as clips that represent individual actions (from the set of actions $\mathcal{A} = \{a_{j}\}_j$), where $\mathcal{S} \cup \mathcal{A} \subseteq \mathcal{C}$~\footnote{In the last expression we have equated the representations of percepts and actions within the clip network with the actions and percepts themselves, in~a slight abuse of notation. In the following, we will be using $s_k$ ($a_k$) to denote the percept (action) clips when the semantics of the clip matters (e.g., whether it is an action or~a percept), and the generic notation $c_k$ when it does not. Formally, there is a distinction between percepts~$s_j$ and actions~$a_j$, and their internal representation (a~memory), usually denoted $\mu(s_j)$ and $\mu(a_j)$, respectively.}. When presented with a percept~$s_i$, the standard PS initiates~a random walk in the clip network, governed by~$P$, and starting from (the clip corresponding to)~$s_i$. The walk is terminated at the first instance an action clip is encountered. This action is then coupled out as~a real action.

This process can be viewed in terms of probability vectors as follows. Each clip $c_i$ can be represented as~a canonical basis vector of an $N$-dimensional real vector space~$\mathcal{V}$, that is, $c_{i}=\left[0,\ldots, 1, 0, \ldots, 0 \right]^{T}$, with the unity at the $i^{th}$ position. The state after one random walk transition is
\begin{align}
    P\,c_{i}  &=\,\sum_{j} p_{ij}\,c_{j}\,,
    \label{eq:Markov chain n-th column classical representation}
\end{align}
which is~a probability vector, i.e., a~vector with real non-negative entries summing to one, representing~a probability distribution over the clip space. This distribution is then sampled from, obtaining some clip $c_{k}$, which, if it represents an action, is coupled out. Otherwise the random walk proceeds from $c_{k}$.

In the spirit of the reinforcement learning paradigm, each round of interaction with the environment is either rewarded or not, and both cases lead to an update of the clip network, by altering the transition probabilities, and/or by altering the clip set itself, which constitutes the \emph{learning aspect} of the PS agent. For an overview of the standard PS model, including examples of update rules, see Ref.~\cite{MautnerMakmalManzanoTierschBriegel2014}.

\subsection{Standard PS with Trapped Ions}\label{subsec:standard PS with trapped ions}

We shall now discuss how the random walk initiated in an standard PS agent can be emulated in a~quantum system, in particular, using laser pulses on~a string of trapped ions. Although~a quantum implementation is not strictly required for the classical random walk of the standard PS agent, such~a construction is the prerequisite for the fully quantized RPS agent that we will discuss in Section~\ref{sec:reflecting projective simulators in ion traps}. For the construction of~a quantum mechanical analogue of the transition matrix~$P$ we start by promoting the real vector space~$\mathcal{V}$ to~a complex Hilbert space~$\mathcal{H}$\,, and representing the clips~$c_{i}$ as orthonormal basis states~$\ket{c_{i}}\,$. We then construct~a unitary~$U_{i}\,$, such that for~a fixed basis state denoted $\ket{0}$ \textemdash this may correspond to some clip state ~$\ket{c_{l}}$ but the particular choice of this fixed state is unimportant \textemdash the components of the state~$U_{i}\ket{0}$ with respect to the clip basis encode the transition amplitudes as dictated by the transition matrix $P$, i.e.,
\begin{align}
    U_{i}\,\ket{0}  &=\,\sum\limits_{j=1}^{N}\,\sqrt{p_{ji}}\,\ket{c_{j}}\,.
    \label{eq:Markov chain n-th column quantum representation}
\end{align}

We can see that~a measurement of the state above in the clip basis recovers the right-hand side of the classical Eq.~(\ref{eq:Markov chain n-th column classical representation}). However~a single unitary cannot encode all the transitions of~$P$. This can be seen quite simply, by noting that the columns of the matrix representation of~$U_{i}$ are required to be orthogonal, while the columns of~$P$ may even be identical. In general, one therefore requires~$N$ distinct unitaries~$U_{i}$ to represent all transitions of~$P$ on an~$N$-dimensional Hilbert space. In other words, the first column, corresponding to the basis state~$\ket{0}$, of the unitary~$U_{i}$ determines the transition probabilities from the clip~$c_{i}$ to any other clip in the sense of Eq.~(\ref{eq:Markov chain n-th column quantum representation}).
Eq.~(\ref{eq:Markov chain n-th column classical representation}) could be recovered even if the amplitudes in Eq.~(\ref{eq:Markov chain n-th column quantum representation}) had arbitrary relative complex phases. These phases are irrelevant in the context of the classical agent, but
for the purpose of the extension to the quantum RPS we restrict the entries of the first column of~$U_{i}$ to be real and positive.

Note that, given the set of unitaries $\{ U_i\}_{i=1}^{N},$ each corresponding to a column of an $N$-state transition matrix~$P$, one can emulate any classical random walk by iterating the measurement of the quantum register (in the clip-basis), resetting the register to the state $\ket{0}$, and applying the~$U_{i}$ corresponding to the prior measurement result.
The capacity to generate such unitaries will, in the next section, be used as~a primitive to construct coherent quantum walks. Here we first analyze how such unitaries can be realized in an ionic set-up.

To proceed, we wish to encode the clip basis in the internal states of~a chain of trapped ions, and the unitaries~$U_{i}$ in the laser pulses driving the transitions between them. We will consider~a setup as described, e.g., in Refs.~\cite{Schindler-Blatt2013,Barreiro-Blatt2011}. A~string of~$^{40}$Ca$^{+}$ ions is confined by~a quadrupole trap (Paul trap). The ion confinement can be described by harmonic potentials, and the Coulomb repulsion of the ions couples the harmonic oscillators, such that the motion of the ions can be captured in terms of their collective normal modes. For each ion, two Zeeman sub-levels, for instance, $\ket{g}:=\ket{S_{1/2,-1/2}}$ and $\ket{e}:=\ket{D_{5/2,-1/2}}$, which can be coupled by~a quadrupole transition, are used to represent the computational basis states of~a single qubit. In turn, we employ the state space of~$k$ qubits as~a representation of the clip network. Hence, the PS implementation we propose requires~$k = \lceil \log_{2}(N) \rceil$ ions for~a network of~$N$ clips.

The required unitaries can be realized with two laser beams~\cite{Schindler-Blatt2013,Barreiro-Blatt2011}, one of which is~a broad beam that is nearly collinear to the ion chain, such that all ions are illuminated. The second laser beam can be focussed to address each ion individually. When operated resonantly at the frequency~$\omega$ corresponding to the transition~$\ket{g}\leftrightarrow\ket{e}$, the first laser laser realizes the collective gate
\begin{align}
    U_{\!X}(\theta)   &=\,\exp\bigl(-i\tfrac{\theta}{2}\sum\limits_{i=1}^{k}X_{i}\bigr)\,,
    \label{eq:collective X rotation gate}
\end{align}
where we use the shorthand notation~$X_{i}$ for \mbox{$\mathds{1}_{1}\otimes\ldots\otimes X_{i}\otimes\ldots\otimes\mathds{1}_{k}\,$}, i.e., the Pauli~$X$ operator for the $i$-th qubit.
The second laser, on the other hand, is applied off-resonance to provide the single-qubit gate
\begin{align}
    U_{\!Z_{i}}(\theta)   &=\,\exp\bigl(-i\tfrac{\theta}{2}Z_{i}\bigr)\,.
    \label{eq:single qubit Z gate}
\end{align}
The operations of Eqs.(\ref{eq:collective X rotation gate}) and~(\ref{eq:single qubit Z gate}) can further be complemented with an entangling gate, such as the Cirac-Zoller~\cite{CiracZoller1995} or M{\o}lmer\textendash S{\o}rensen~\cite{MoelmerSoerensen1999} gate, to form~a universal set of quantum gates, and hence provide the possibility to construct the unitaries~$U_{i}$ in principle. In general, the aim is to determine~a sequence of operations with~$(N-1)$ free parameters~$\theta_{1},\ldots,\theta_{N-1}\,$, such that all entries of the first column of the resulting overall unitary~$U(\theta_{1},\ldots,\theta_{N-1})$ are real and positive, and for appropriate choices of the~$\theta_{j}$ their squares can form any arbitrary probability distribution~$\{p_{n}\}_{n=1}^{N}\,$, with~$\sum_{n}p_{n}=1\,$. The freedom in the choice of parameters allows for all of the operators~$U_{i}$ to be represented by some specific choices of the~$\theta_{j}\,$. In particular, the agent is considered to operate based on~a fixed internal architecture, in particular the tuning of the angles should have~a simple operational meaning. At every step of the learning process, the agent only updates~a set of parameters, here the~$\theta_{i}\,$, corresponding to the duration of some laser pulses within~a fixed sequence. For instance, in the very simple case of~a clip network with only two clips, the required unitary can be chosen to be~a Pauli-$Y$ rotation of~a single qubit, given by
\begin{align}
    U_{\hspace*{-0.5pt}Y}(\theta) &=\,\exp\bigl(-i\,\tfrac{\theta}{2}\,Y\bigr)\,=\,
        \begin{pmatrix}\cos\tfrac{\theta}{2}  &   -\sin\tfrac{\theta}{2} \\[1mm]
        \sin\tfrac{\theta}{2}  &   \hspace*{2mm} \cos\tfrac{\theta}{2} \end{pmatrix}\,,
    \label{eq:single qubit Y rotation}
\end{align}
which can be realized by three laser pulses, i.e.,
\begin{align}
    U_{\hspace*{-0.5pt}Y_{\hspace*{-0.5pt}j}}(\theta) &=\,U_{\!X}(-\tfrac{\pi}{2})\,U_{\!Z_{j}}(\theta)\,U_{\!X}(\tfrac{\pi}{2})\,,
    \label{eq:single qubit Y rotation decomposition}
\end{align}
and where we have included the qubit label~$j$ for later convenience.

As we have mentioned earlier, the `probability unitaries' presented above will become the building blocks of the quantum PS agent. The second, and last, crucial ingredient in our construction is the technique of adding coherent control, which we shall briefly present next.

\subsection{Coherent Controlization}\label{subsec:Coherent Controlization}
Adding coherent control entails coherently conditioning (unitary) operations on the state of~a control system.
More formally, this is represented as~a mapping from a~set of unitaries $\{U_i\}_{i=1}^{M}$, acting on a Hilbert space $\mathcal{H}$, to~a single controlled unitary~$U$ of the form
\begin{eqnarray}
U \ket{j} \otimes \ket{\psi} = \ket{j} \otimes U_j \ket{\psi},\ \forall \ket{\psi},
\end{eqnarray}
which acts on $\mathcal{H}_{C}\otimes\mathcal{H}$\,, where~$\mathcal{H}_{C}$ is an (at least) $M$\textendash dimensional Hilbert space, and~$\{\ket{j}\}$ is an orthonormal basis of~$\mathcal{H}_{C}$\,. Practically, this mapping may be understood as~a physical procedure of adding quantum control to individual elementary operations~\cite{FriisDunjkoDuerBriegel2014}. We refer to such mappings and the associated physical processes, which implicitly feature in many quantum algorithms~\cite{Shor1994,Kitaev1996}, as \emph{coherent controlization}. As we will discuss in Section~\ref{sec:reflecting projective simulators in ion traps}, coherent controlization forms an essential part of the construction of the quantum RPS agent.

As~a first instance of its applicability, coherent controlization provides an elegant method to generically assemble and combine probability unitaries. The latter may also be assembled in other, sometimes more efficient ways, and one alternative construction is provided in the \hyperref[Appendix]{Appendix}. Nonetheless, the construction of the probability unitaries using coherent controlization offers the opportunity to illustrate this method on~a simple and useful example.

Before we begin, let us recall the task at hand. For~a given probability distribution~$\{p_{ij}\}_{i=1}^{N}$, corresponding to the~$j$-th column of the stochastic matrix~$P$, we wish to construct the associated unitary~$U_{\!j}$, such that the first column of~$U_{\!j}$ has real and positive entries~$\sqrt{p_{ij}}$, with $i=1,\ldots,N$.

As the elementary operations that depend on these parameters we select single-qubit $Y$~rotations~$U_{Y}(\theta_{i})$, which, for~a trapped ion setup, may be realized as in Eq.~(\ref{eq:single qubit Y rotation decomposition}), and where we drop the label~$Y$ for ease of notation. Any probability unitary~$U(\theta_{1},\ldots,\theta_{N-1})$ on an $N$-clip network can then be assembled by~a nested scheme of coherent controlization on~$k$ qubits, where~$k$ is the smallest integer that is larger than~$\log_{2}(N)$. For simplicity, let us assume here that the size of the clip network is such that $\mathbb{N}\ni\log_{2}(N)=k$, which can always be achieved by duplicating some clips.

\begin{figure}[ht!]
\includegraphics[width=0.49\textwidth]{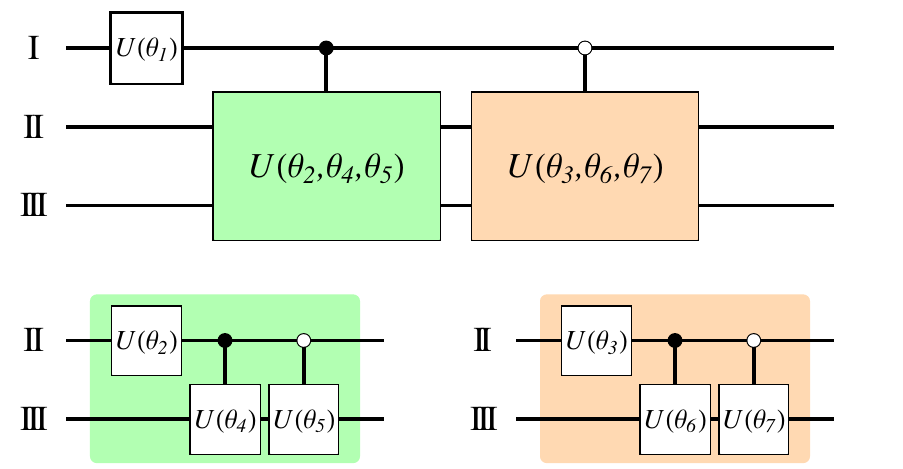}
\caption{\label{fig:nesting scheme}
\textbf{Coherent controlization}. The circuit diagrams show the construction of~a three-qubit probability unitary~(\ref{eq:three qubit coherent controlization}), using coherent controlization. The filled dots ``$\bullet$" on the controlled operations indicate that the unitaries on the target are conditioned on the control qubit state~$\ket{0}$, while the hollow dots ``$\circ$" represent conditioning on the control qubit state~$\ket{1}$.}
\end{figure}

For~a two-clip probability distribution~$\{p_{1},p_{2}\}$, the probability unitary is trivially realized by~a single-qubit $Y$ rotation~$U(\theta_{1})$, with $p_{1}=\cos^{2}(\theta_{1}/2)$ and $p_{2}=\sin^{2}(\theta_{1}/2)$. To extend this to~a four-clip probability unitary $U(\theta_{1},\theta_{2},\theta{3})$, with probability distribution $\{p_{1}\pr,p_{2}\pr,p_{3}\pr,p_{4}\pr\}$, one adds~a second qubit, hence~$k=2$, and starts again with the operation~$U(\theta_{1})$ on the first qubit, where~$p_{1}=p_{1}\pr+p_{2}\pr$ and~$p_{2}=p_{3}\pr+p_{4}\pr$. This is followed by two controlled $Y$ rotations of the second qubit, conditioned on the state of the first, that is,~$U(\theta_{2})$ is applied if the first qubit is in the state~$\ket{0}$, while~$U(\theta_{3})$ is applied when the first qubit is in the state~$\ket{1}$. The corresponding angles are determined from the renormalized probabilities within the respective subspaces, i.e., $\cos^{2}(\theta_{2}/2)=p_{1}\pr/(p_{1}\pr+p_{2}\pr)$ and $\cos^{2}(\theta_{3}/2)=p_{3}\pr/(p_{3}\pr+p_{4}\pr)$.

For larger values of~$k$, the controlization becomes nested, see Fig.~\ref{fig:nesting scheme}, e.g., for $k=3$ ($N=8$), the lowest level of single qubit operations, here~$U(\theta_{2})$ and~$U(\theta_{3})$, is followed by controlled operations on~a third qubit. Labeling the qubits as~$\mathrm{I}$,~$\mathrm{I\hspace*{-1pt}I}$, and~$\mathrm{I\hspace*{-1pt}I\hspace*{-1pt}I}$, we may write the corresponding probability unitary as
\begin{align}
    U(\theta_{1},\ldots,\theta_{7}) &=\,
        \bigl[U(\theta_{2},\theta_{4},\theta_{5})\oplus U(\theta_{3},\theta_{6},\theta_{7})\bigr]_{\raisebox{0.6pt}{\scriptsize{$\mathrm{I},\mathrm{I\hspace*{-1pt}I},\mathrm{I\hspace*{-1pt}I\hspace*{-1pt}I}$}}}
        \nonumber\\
        &\ \ \times\bigl[U(\theta_{1})_{\raisebox{-2pt}{\scriptsize{$\mathrm{I}$}}}\otimes \mathds{1}_{\raisebox{-2pt}{\scriptsize{$\mathrm{I\hspace*{-1pt}I},\mathrm{I\hspace*{-1pt}I\hspace*{-1pt}I}$}}}\bigr]\,,
        \label{eq:three qubit coherent controlization level 1}
\end{align}
where the controlled two-qubit operations are given by
\begin{subequations}
\label{eq:three qubit coherent controlization}
\begin{align}
    U(\theta_{2},\theta_{4},\theta_{5}) &=\,
        \bigl[U(\theta_{4})\oplus U(\theta_{5})\bigr]_{\raisebox{0.6pt}{\scriptsize{$\mathrm{I\hspace*{-1pt}I},\mathrm{I\hspace*{-1pt}I\hspace*{-1pt}I}$}}}
        \bigl[U(\theta_{2})_{\raisebox{-2pt}{\scriptsize{$\mathrm{I\hspace*{-1pt}I}$}}}\otimes\mathds{1}_{\raisebox{-2pt}{\scriptsize{$\mathrm{I\hspace*{-1pt}I\hspace*{-1pt}I}$}}}\bigr]\,,
        \label{eq:three qubit coherent controlization level 2 1}\\
    U(\theta_{3},\theta_{6},\theta_{7}) &=\,
        \bigl[U(\theta_{6})\oplus U(\theta_{7})\bigr]_{\raisebox{0.6pt}{\scriptsize{$\mathrm{I\hspace*{-1pt}I},\mathrm{I\hspace*{-1pt}I\hspace*{-1pt}I}$}}}
        \bigl[U(\theta_{3})_{\raisebox{-2pt}{\scriptsize{$\mathrm{I\hspace*{-1pt}I}$}}}\otimes
        \mathds{1}_{\raisebox{-2pt}{\scriptsize{$\mathrm{I\hspace*{-1pt}I\hspace*{-1pt}I}$}}}\bigr]\,.
        \label{eq:three qubit coherent controlization level 2 2}
\end{align}
\end{subequations}

As we have argued above, coherent controlization allows for the construction of general probability unitaries from basic single-qubit probability unitaries. Despite the simple appearance of the circuits in Fig.~\ref{fig:nesting scheme}, the practical implementation of coherent controlization requires additional attention. In fact, it is generally impossible to decompose quantum-controlled operations $\operatorname{ctrl}(U)$ into individual gates~$\operatorname{ctrl}(U)=G_{1}\,U\,G_{2}$, such that the~$G_{i}$ are independent of~$U$, which implies that the gates~$G_{i}$ may not be specified if~$U$ is unknown~\cite{AraujoFeixCostaBrukner2013,ThompsonGuModiVedral2013}. This seems to suggest that coherent controlization requires computational effort in its implementation. However, for the ionic implementation that we will discuss next, we exploit additional degrees of freedom of the physical setup to perform coherent controlization in~a generic way.

\begin{figure}[ht!]
\includegraphics[width=0.4\textwidth]{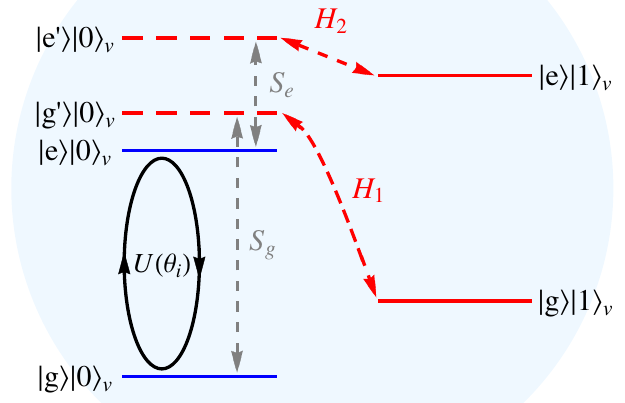}
\caption{\label{fig:control with ions}
\textbf{Level structure of trapped ions}. An illustration of the energy levels of one of the ions in the trap is shown. Two levels,~$\ket{g}$ and~$\ket{e}$, are chosen to represent~the qubit, while the auxiliary levels~$\ket{g\pr}$ and~$\ket{e\pr}$, and the first excited state~$\ket{1}_{v}$ of the common vibrational mode are used in the process of coherent controlization. The transitions indicated by~$H_{1}$, $H_{2}$, $S_{g}$, and~$S_{e}$ can be realized by appropriately detuned $Y$-pulses.}
\end{figure}

\subsection{Coherent Controlization in Trapped Ions}\label{subsec:Coherent Controlization in Trapped Ions}
We shall now discuss how quantum control can be practically added to unitaries that are realized by laser pulses in~a trapped ion setup, based on the scheme introduced in Ref.~\cite{FriisDunjkoDuerBriegel2014}. As an example we give the explicit pulse decomposition that realizes the two-qubit unitary~$U(\theta_{1},\theta_{2},\theta_{3})$, which can be viewed as~a special case of Eq.~(\ref{eq:three qubit coherent controlization level 2 1}) for $\theta_{4},\ldots,\theta_{7}=0$, where we use two ions, labeled~$\mathrm{I}$ and~$\mathrm{I\hspace*{-1pt}I}$, respectively, before we explain how this method is generalized to the control of $k$-qubit unitaries.

To start, we note that the operation $\bigl[U(\theta_{1})_{\raisebox{-2pt}{\scriptsize{$\mathrm{I}$}}}\otimes\mathds{1}_{\raisebox{-2pt}{\scriptsize{$\mathrm{I\hspace*{-1pt}I}$}}}\bigr]$ can be trivially implemented by the pulse sequence of Eq.~(\ref{eq:single qubit Y rotation decomposition}), and we can thus focus our attention on the remaining term $\bigl[U(\theta_{2})\oplus U(\theta_{3})\bigr]_{\raisebox{0.6pt}{\scriptsize{$\mathrm{I},\mathrm{I\hspace*{-1pt}I}$}}}$. Apart from the laser pulses for the elementary operations~$U(\theta_{2})$ and~$U(\theta_{3})$, our scheme for their coherent controlization also consists of a~number of additional~$Y$ rotations in $2$-dimensional subspaces of the ionic energy levels other than the one spanned by~$\ket{g}$ and~$\ket{e}$, see Fig.~\ref{fig:control with ions}. We will use additional superscripts, e.g.,~$U_{\!Y_{i}}^{\raisebox{0pt}{\tiny{\#}}}\,$, where the labels~``$\#$" identify different detuning frequencies, and the subscript~$i\in\{\mathrm{I},\mathrm{I\hspace*{-1pt}I}\}$ identifies the ion, to distinguish these operations. Furthermore, we make use of one of the common vibrational modes, which we assume has been cooled to the ground state~$\ket{0}_{v}\,$, before the following steps are executed.

\begin{enumerate}[\hspace*{1.5mm}(i)]
\item{\label{item:adding control i}
    Cirac-Zoller~\cite{CiracZoller1995,SchmidtKaler-Blatt2003} method:\ A~sequence of appropriately blue-detuned laser pulses is applied on ion~$\mathrm{I}$ to realize~$U_{\!Y_{\mathrm{I}}}^{\raisebox{0pt}{\tiny{$C\!Z$}}}(\pi)$, which transfers the population of~$\ket{g}_{\mathrm{I}}\!\ket{0}_{\mathrm{v}}$ to~$\ket{e}_{\mathrm{I}}\!\ket{1}_{\mathrm{v}}\,$. This step encodes the state of qubit~$\mathrm{I}$ in the vibrational mode, i.e., the initial state of the form $\bigl(\alpha\ket{g}_{\mathrm{I}}+\beta\ket{e}_{\mathrm{I}}\bigr)\ket{\psi}_{\mathrm{I\hspace*{-1pt}I}}\ket{0}_{v}$ is transformed to $\ket{e}_{\mathrm{I}}\ket{\psi}_{\mathrm{I\hspace*{-1pt}I}}\bigl(\beta\ket{0}_{v}+\alpha\ket{1}_{v}\bigr)$.}
\item{\label{item:adding control ii}
    Hiding:\ Red-detuned laser pulses corresponding to~$U_{\!Y_{\mathrm{I\hspace*{-1pt}I}}}^{\raisebox{0pt}{\tiny{$H_{1}$}}}(\pi)$ and~$U_{\!Y_{\mathrm{I\hspace*{-1pt}I}}}^{\raisebox{0pt}{\tiny{$H_{2}$}}}(\pi)$ are applied to ion~$\mathrm{I\hspace*{-1pt}I}$ to transfer the populations from~$\ket{g}_{\mathrm{I\hspace*{-1pt}I}}\!\ket{1}_{v}$ to~$\ket{g\pr}_{\mathrm{I\hspace*{-1pt}I}}\!\ket{0}_{v}$\,, as well as from~$\ket{e}_{\mathrm{I\hspace*{-1pt}I}}\!\ket{1}_{v}$ to~$\ket{e\pr}_{\mathrm{I\hspace*{-1pt}I}}\!\ket{0}_{v}\,$, as illustrated in Fig.~\ref{fig:control with ions}. Denoting the state~$\psi$ encoded in the levels~$\ket{g\pr}_{\mathrm{I\hspace*{-1pt}I}}$ and~$\ket{e\pr}_{\mathrm{I\hspace*{-1pt}I}}$ as~$\ket{\psi\pr}_{\mathrm{I\hspace*{-1pt}I}}\,$, we may write the overall state after this step as $\ket{e}_{\mathrm{I}}\bigl(\alpha\ket{\psi\pr}_{\mathrm{I\hspace*{-1pt}I}}+\beta\ket{\psi}_{\mathrm{I\hspace*{-1pt}I}}\bigr)\ket{0}_{v}$.}
\item{\label{item:adding control iii}
    $U(\theta_{3})$:\ The pulse sequence that realizes~$U(\theta_{3})$ is applied to ion~$\mathrm{I\hspace*{-1pt}I}$, which leaves the system in the state
    $\ket{e}_{\mathrm{I}}\bigl(\alpha\ket{\psi\pr}_{\mathrm{I\hspace*{-1pt}I}}+
    \beta\,U(\theta_{3})\ket{\psi}_{\mathrm{I\hspace*{-1pt}I}}\bigr)\ket{0}_{v}$.}
\item{\label{item:adding control iv}
    Switching:\ To exchange the primed and unprimed levels, laser pulses for~$U_{\!Y_{\mathrm{I\hspace*{-1pt}I}}}^{\raisebox{-1pt}{\tiny{$S_{g}$}}}(\pi)$ and~$U_{\!Y_{\mathrm{I\hspace*{-1pt}I}}}^{\raisebox{-1pt}{\tiny{$S_{e}$}}}(\pi)$, which are blue- and red-detuned, respectively, are applied to ion~$\mathrm{I\hspace*{-1pt}I}$, see Fig.~\ref{fig:control with ions}. The resulting overall state after these operations is
    $\ket{e}_{\mathrm{I}}\bigl(\alpha\,\ket{\psi}_{\mathrm{I\hspace*{-1pt}I}}+
    \beta\ket{(U(\theta_{3})\psi)\pr}_{\mathrm{I\hspace*{-1pt}I}}\bigr)\ket{0}_{v}$.}
\item{\label{item:adding control v}
    $U(\theta_{2})$:\ The pulse sequence that realizes~$U(\theta_{2})$ is applied to ion~$\mathrm{I\hspace*{-1pt}I}$, such that the system is now in the state
    $\ket{e}_{\mathrm{I}}\bigl(\alpha\,U(\theta_{2})\ket{\psi}_{\mathrm{I\hspace*{-1pt}I}}+
    \beta\ket{(U(\theta_{3})\psi)\pr}_{\mathrm{I\hspace*{-1pt}I}}\bigr)\ket{0}_{v}$.}
\item{\label{item:adding control vinew}
    Switching:\ The primed and unprimed levels are exchanged again using the laser pulses for~$U_{\!Y_{\mathrm{I\hspace*{-1pt}I}}}^{\raisebox{-1pt}{\tiny{$S_{g}$}}}(\pi)$ and~$U_{\!Y_{\mathrm{I\hspace*{-1pt}I}}}^{\raisebox{-1pt}{\tiny{$S_{e}$}}}(\pi)$ on ion~$\mathrm{I\hspace*{-1pt}I}$, which leaves the system in the state
    $\ket{e}_{\mathrm{I}}\bigl(\alpha\,\ket{(U(\theta_{2})\psi)\pr}_{\mathrm{I\hspace*{-1pt}I}}+
    \beta\,U(\theta_{3})\ket{\psi}_{\mathrm{I\hspace*{-1pt}I}}\bigr)\ket{0}_{v}$.}
\item{\label{item:adding control vi}
    Unhiding:\ The hiding operations of step~(\ref{item:adding control ii}) are reversed by the application of~$U_{\!Y_{\mathrm{I\hspace*{-1pt}I}}}^{\raisebox{0pt}{\tiny{$H_{1}$}}}(-\pi)$ and~$U_{\!Y_{\mathrm{I\hspace*{-1pt}I}}}^{\raisebox{0pt}{\tiny{$H_{2}$}}}(-\pi)$ to ion~$\mathrm{I\hspace*{-1pt}I}$, leaving the system in the state
    $\ket{e}_{\mathrm{I}}\bigl(\alpha\,U(\theta_{2})\ket{\psi}_{\mathrm{I\hspace*{-1pt}I}}\ket{1}_{v}+
    \beta\,U(\theta_{3})\ket{\psi}_{\mathrm{I\hspace*{-1pt}I}}\ket{0}_{v}\bigr)$.}
\item{\label{item:adding control vii}
    Return control:\ Finally,~$U_{\!Y_{\mathrm{I}}}^{\raisebox{0pt}{\tiny{$C\!Z$}}}(-\pi)$ is applied to ion~$\mathrm{I}$, which returns the control from the vibrational mode, and~a provides the desired state
    $\bigl(\alpha\,\ket{g}_{\mathrm{I}}U(\theta_{2})\ket{\psi}_{\mathrm{I\hspace*{-1pt}I}}+
    \beta\,\ket{e}_{\mathrm{I}}U(\theta_{3})\ket{\psi}_{\mathrm{I\hspace*{-1pt}I}}\bigr)\ket{0}_{v}$, that is, the unitary $U(\theta_{2})$ acts on ion~$\mathrm{I\hspace*{-1pt}I}$, when ion~$\mathrm{I}$ is in the state~$\ket{g}_{\mathrm{I}}$, while $U(\theta_{3})$ acts upon the subspace in which the first ion is in the state~$\ket{e}_{\mathrm{I}}$.}
\end{enumerate}
If required, the scheme laid out in steps~(\ref{item:adding control i})-(\ref{item:adding control vii}) may be straightforwardly extended to larger clip spaces by increasing the number of control qubits and vibrational modes used. Each $Y$ rotation in principle requires~$3$ individual pulses, see Eq.~(\ref{eq:single qubit Y rotation decomposition}), but the collective $X$ rotations for the operations~$U(\theta_{i})$ can be subsumed into two single pulses~$U_{\!X}(\tfrac{\pi}{2})$ and~$U_{\!X}(-\tfrac{\pi}{2})$ at the start and at the end of the entire pulse sequence, respectively. We hence find that the overall number of elementary laser pulses necessary to assemble~a $k$-qubit probability unitary is given by $(7\times2^{k+2}-24k-29)$ for $k\geq2$. Note that an exponential scaling in terms of the qubits used is inevitable, as $k$ qubits encode $2^k$ probabilities, and we must have the freedom to specify each one of these. In terms of the state space of the ECM network (clip number) the scaling is linear.

In such~a process $(k-1)$ vibrational modes of different frequencies are used to generalize steps~(\ref{item:adding control i}) and~(\ref{item:adding control vii}) to condition $(k-1)$-qubit operations on the state of the first qubit, i.e., by transferring the populations (exclusively) between $\ket{g}_{\mathrm{I}}\ket{0\ldots0}_{v_{1},\ldots,v_{k-1}}$ and $\ket{e}_{\mathrm{I}}\ket{1\ldots1}_{v_{1},\ldots,v_{k-1}}\,$.

Next, we give the basics of the classical and quantum RPS agent models, and show how the two components\textemdash coherent controlization and probability unitaries\textemdash can be utilized to construct these in systems of trapped ions.

\section{Reflecting PS with Trapped Ions}\label{sec:reflecting projective simulators in ion traps}

We now turn to the so-called reflecting projective simulation (RPS) agent introduced in Ref.~\cite{PaparoDunjkoMakmalMartinDelgadoBriegel2014}. The central aim of the RPS is to output the actions according to~a specific distribution, which we shall specify shortly, that is updated, indirectly, as the ECM network is modified throughout the learning process. Here, the clip network~$\mathcal{C}$ is disjoint, and it comprises unconnected percept-specific subnetworks with associated stochastic (ergodic and time-reversible) matrices~$P_{k}=[(p_{k})_{ij}]$, for each percept~$s_{k}\,$.

Depending on which percept is observed, the random walk is executed on the corresponding percept-specific (sub-)network, where it is continued until the Markov chain~$P_{k}$ is (approximately) mixed, that is, until the respective stationary distribution~$\piP{P_{k}}$, which has support over the entire clip space, is (approximately) reached. The agent then samples from the obtained distribution, and iterates the procedure (which requires re-mixing of the Markov chain) until an action is hit.
More specifically, the RPS agent is designed to output (a good approximation) of the \emph{tailed distribution} $\pitildeP{P_{k}}$ defined as
\begin{align}
    (\pitildeP{P_{k}})_{j} &=\,
    \begin{cases}\displaystyle
        \mathcal{N}\times(\piP{P_{k}})_{j}\,, &    \mbox{if $c_{j}$ is an action,}\\
        0\,,   &   \mbox{otherwise,}
    \end{cases}
\end{align}
where $\mathcal{N}$ is~a normalization factor such that~\mbox{$\sum_{j}(\pitildeP{P_{k}})_{j}=1$}. That is, the re-normalized distribution $\piP{P_{k}}$ truncated such that it has support only over the action space.

Despite the differences in the walk termination criteria of the standard PS and RPS models, all the operational elements required for an emulation of~a classical RPS agent in an ionic set-up have already been presented in the last section, as the previously described construction enables the emulation of any classical random walk.

In the remainder of this section, we aim to show how the \emph{quantum} RPS agent, which employs~a truly coherent quantum walk (in the sense of \cite{Szegedy2004,MagniezNayakRolandSantha2011}) to obtain~a quadratic speed-up over the classical RPS agent, can be implemented based on the coherent controlization of unitaries as discussed in Section~\ref{subsec:Coherent Controlization in Trapped Ions}. For notational simplicity, we will from this point on ignore the subscript~$k$ indicating the percept the network in question corresponds to, unless it is specifically required.

The central process of the quantum RPS model, the basics of which we present next, is~a so-called Szegedy-type quantum random walk, see, e.g., Ref.~\cite{MagniezNayakRolandSantha2011}, that is performed on the percept-specific ECM (sub-)network. These Szegedy-type quantum random walks are used in the quantum RPS agent in order to output an action distributed according to the tailed stationary distribution $\pitildeP{P}$ with~a quadratically decreased number of elementary diffusion steps, as compared to~a classical RPS agent.

As the structure of this decision-making process is rather involved, let us briefly sketch it out here, before proceeding in more detail. The basic building block of~a Szegedy-type walk, is the elementary \emph{diffusion unitary} $U_{\!P}$, which acts on~a two register system, each one of sufficient dimensionality to represent the entire clip network. One application of $U_{\!P}$ can be considered as the analog of one step of the classical walk governed by the transition matrix~$P$. The Szegedy walk operator~$W(P)$, on the other hand, is constructed using four applications of $U_{\!P}$ (or its inverse), and some quantum operations which are independent from~$P$. One of the distinct properties of the operator~$W(P)$ is that its unique~$(+1)$ eigenstate~$\ket{\piP{P}\pr}$ is~a particular coherent encoding of the stationary distribution~$\piP{P}$ of the Markov chain. Exploiting this property, and using~a modified Kitaev phase estimation algorithm~\cite{Kitaev1996}, we can construct an approximate reflection operator (ARO), which reflects over the state $\ket{\piP{P}\pr}$. The speed-up achieved in the quantum RPS originates, in part, from the efficiency of the construction of the ARO operator in terms of the number of applications of the diffusion unitary $U_{\!P}$, relative to the mixing time of the Markov chain as specified by~$P$.

The ARO operator above can then be used in search algorithms (e.g., as in Refs.~\cite{Szegedy2004,MagniezNayakRolandSantha2011}), as well as in the decision-making process of the RPS agent, which can be seen as~a Grover-type~\cite{Grover1996} reflection process in the following sense. Upon the system, initialized in the state $\ket{\piP{P}\pr}$, one sequentially applies a `check' operator, which adds~a relative phase of $(-1)$ to all basis states corresponding to actions, followed by the ARO operator, which reflects over the coherent encoding of the stationary distribution. This, like in the Grover algorithm, induces~a sequence of rotations in a $2$-dimensional workspace, which, after a certain number of iterations, guarantees that the system state has~a constant overlap with the state encoding the aforementioned tailed distribution. The second component of the quantum speed-up lies in the number of these iterations, which inherits the quadratic improvement that is characteristic to Grover's algorithm. With this in mind, let us now give further details of the building blocks of the quantum RPS.

\subsection{The Szegedy Walk Operator}\label{subsec:The Szegedy Walk operator}

As we have argued previously,~a unitary on an $N$-dimensional Hilbert space is not capable of representing all transitions of an arbitrary Markov chain over~a network of~$N$ clips. For this reason, the classical random walk for~a given transition matrix~$P$ that we have described in Section~\ref{subsec:standard PS with trapped ions} is realized by, in general,~$N$ unitaries~$U_{1}\,,~\ldots,~U_{N}\,$, where~$U_{i}$ is associated with the \mbox{$i$-th} column of~$P$. In the Szegedy-type approach to quantum

\begin{figure}[ht!]
(a)
\includegraphics[width=0.42\textwidth]{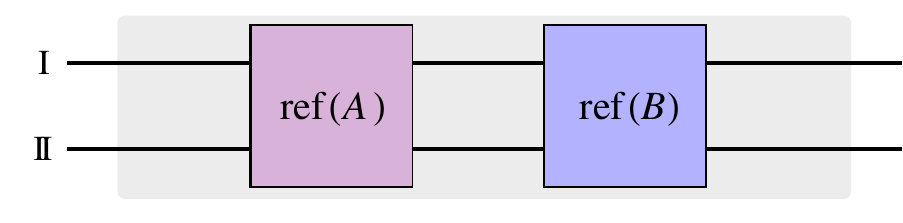}\\
(b)
\includegraphics[width=0.42\textwidth]{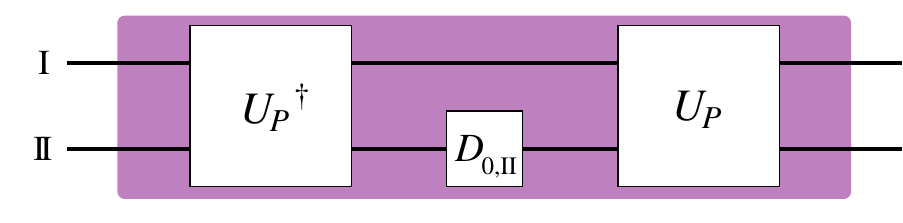}\\
(c)
\includegraphics[width=0.42\textwidth]{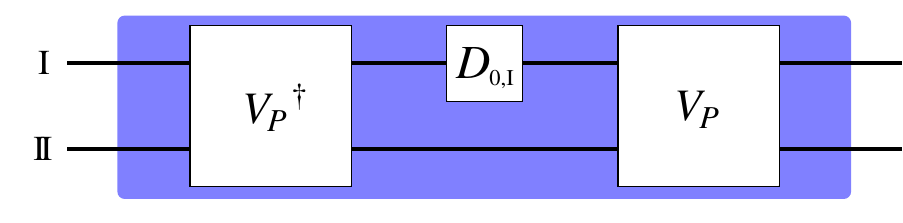}
\caption{\label{fig:generalized walk operator}
\textbf{Szegedy walk operator}.
The circuit representations of the Szegedy walk operator~$W(P)$ of Eq.~(\ref{eq:refB refA}), as well as the reflections over~$A$ and~$B$ [see Eq.~(\ref{eq:spaces A and B})] are shown in~Fig.~\ref{fig:generalized walk operator}~(a), (b), and (c), respectively. The reflection over~$A$~($B$) is fully determined by the walk operator~$U_{\!P}$~($V_{\!P}$) and~a reflection over~$\ket{0}\,$, i.e., $D_{0}=2\ket{0}\!\!\bra{0}-\mathds{1}_{\!N}$\,.}
\end{figure}
\noindent
walks, two copies,~$\mathcal{H}_{\mathrm{I}}$ and $\mathcal{H}_{\mathrm{I\hspace*{-1pt}I}}\,$, of an~$N$-dimensional Hilbert space, i.e., $\dim(\mathcal{H}_{\mathrm{I}})=\dim(\mathcal{H}_{\mathrm{I\hspace*{-1pt}I}})=N\,$ are used to accommodate for all the required degrees of freedom. For~a time-reversible Markov chain we define the unitary walk operators~$U_{\!P}$ and~$V_{\!P}$ as
\begin{subequations}
\label{eq:Szegedy walk operators}
\begin{align}
    U_{\!P}\,\ket{c_{i}}_{\mathrm{I}}\,\ket{0}_{\mathrm{I\hspace*{-1pt}I}}  &=\,\ket{c_{i}}_{\mathrm{I}}\,U_{i}\ket{0}_{\mathrm{I\hspace*{-1pt}I}}\,,
    \label{eq:Szegedy walk operator U}\\[2mm]
    V_{\!P}\,\ket{0}_{\mathrm{I}}\,\ket{c_{i}}_{\mathrm{I\hspace*{-1pt}I}}  &=\,\bigl(U_{i}\ket{0}_{\mathrm{I}}\bigr)\,\ket{c_{i}}_{\mathrm{I\hspace*{-1pt}I}}\,,
    \label{eq:Szegedy walk operator U reversed}
\end{align}
\end{subequations}
where $\{\ket{c_{i}}_{\mathrm{I/I\hspace*{-1pt}I}}|i=1,\ldots,N\}$ form bases of~$\mathcal{H}_{\mathrm{I/II}}\,$. The unitaries~$U_{i}$ act on $\ket{0}_{\mathrm{I/I\hspace*{-1pt}I}}=\ket{c_{1}}_{\mathrm{I/I\hspace*{-1pt}I}}\,$ according to
\begin{align}
    U_{i}\ket{0}_{\mathrm{I/I\hspace*{-1pt}I}}    &=\,\sum_{j=1}^{N}\sqrt{p^{\raisebox{-2pt}{\tiny{$ $}}}_{ji}}\ket{c_{j}}_{\mathrm{I/I\hspace*{-1pt}I}}\,.
    \label{eq:Szegedy walk basic Un}
\end{align}

In the context of quantum RPS agents, we assume that the underlying ergodic Markov chain is time-reversible, i.e., it satisfies detailed balance. Although the Szegedy-type walk can be defined even if this is not the case, one would additionally require access to the time-reversed transition matrix\footnote{Note that the asterisk on the time-reversed transition matrix $P^{*}=(p^{*}_{ij})\,$ does not indicate complex conjugation, and its components are given by $p^{*}_{ij} =p_{ji}\,(\piP{P})_{i}/(\piP{P})_{j}$.} $P^{*}$ in such~a situation. Here, we will present the construction in the most general terms, with the implicit understanding that for the RPS, the unitary $V_{\!P}$ can be obtained from $U_{\!P}$ by swapping the registers prior to, and after the application of~$U_{\!P}$. With the operators $U_{\!P}$ and $V_{\!P}$ at hand, we can now proceed with the construction of the Szegedy walk operator~$W(P)$, which is implemented by reflecting over the spaces~$A$ and~$B$, defined as
\begin{subequations}
\label{eq:spaces A and B}
\begin{align}
A   &:=\,\operatorname{span}\{\ket{\psi_{i}^{\raisebox{-1pt}{\tiny{$A$}}}\!}=U_{\!P}\ket{c_{i}}_{\mathrm{I}}\ket{0}_{\mathrm{I\hspace*{-1pt}I}}\,|\,i=1,\ldots,N\}\,,
    \label{eq:space A}\\[1mm]
B   &:=\,\operatorname{span}\{\ket{\psi_{i}^{\raisebox{-1pt}{\tiny{$B$}}}\!}=V_{\!P}\ket{0}_{\mathrm{I}}\ket{c_{i}}_{\mathrm{I\hspace*{-1pt}I}}\,|\,i=1,\ldots,N\}\,.
    \label{eq:space B}
\end{align}
\end{subequations}

The generalized walk operator is then defined as
\begin{align}
    W(P)   &=\,\operatorname{ref}(B)\,\operatorname{ref}(A)\,,
    \label{eq:refB refA}
\end{align}
where, for~$X=A,\,B\,$, we have
\begin{align}
    \operatorname{ref}(X)   &=\,2\sum\limits_{i=1}^{N}\ket{\psi_{i}^{\raisebox{-1pt}{\tiny{$X$}}}}\!\!\bra{\psi_{i}^{\raisebox{-1pt}{\tiny{$X$}}}}
    \,-\,\mathds{1}_{\!N\times N}\,.
    \label{eq:reflection over X}
\end{align}
The two operators~$\operatorname{ref}(A)$ and~$\operatorname{ref}(B)$ are constructed from the diffusion operators,~$U_{\!P}$ and~$V_{\!P}\,$, along with reflections over~$\ket{0}_{\mathrm{I}}\,$, and~$\ket{0}_{\mathrm{I\hspace*{-1pt}I}}$ denoted~$D_{0,\mathrm{I}}$ and~$D_{0,\mathrm{I\hspace*{-1pt}I}}\,$, respectively, as shown in Fig.~\ref{fig:generalized walk operator}. The unique~$(+1)$ eigenstate~$\ket{\piP{P}\pr}$ of the Szegedy walk operator~$W(P)$, which coherently encodes the stationary distribution~$\piP{P}$ on the two registers, is given by
\begin{align}
    \ket{\piP{P}\pr}    &=\,U_{\!P}\,\ket{\piP{P}}_{\mathrm{I}}\,\ket{0}_{\mathrm{I\hspace*{-1pt}I}}\,=\,
        \sum\limits_{i}\sqrt{(\piP{P})_{i}}\,\ket{c_{i}}_{\mathrm{I}}\,U_{i}\,\ket{0}_{\mathrm{I\hspace*{-1pt}I}}\,.
    \label{eq:stationary state pi pr}
\end{align}

\subsection{The Approximate Reflection Operator}\label{subsec:The approx refl operator}

\begin{figure}[ht!]
(a)\includegraphics[width=0.42\textwidth]{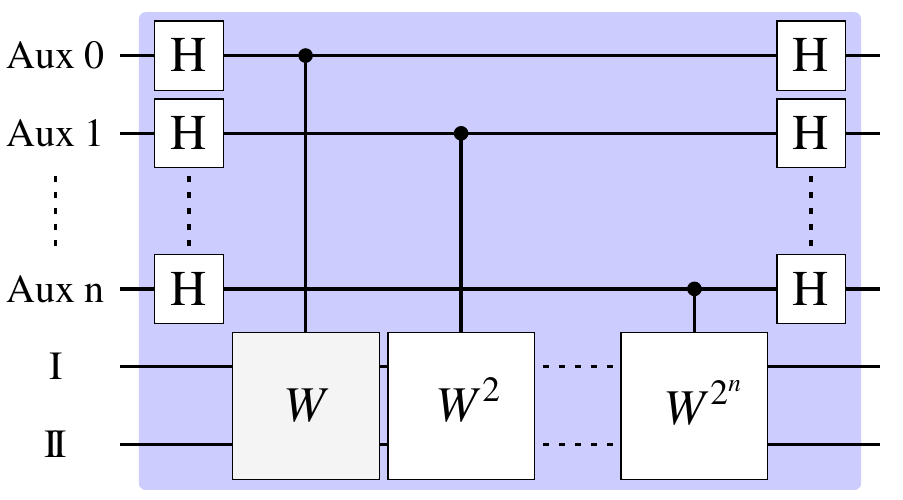}
(b)\includegraphics[width=0.42\textwidth]{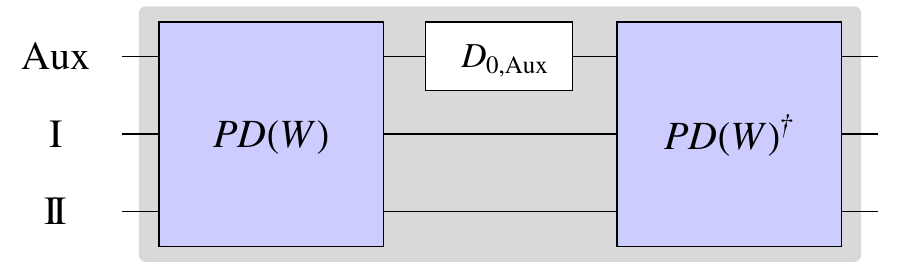}
\caption{\label{fig:phase detection circuit}
\textbf{Phase detection and approximate reflection}. The circuit in Fig.~\ref{fig:phase detection circuit}~(a) shows the phase detection operator~$P\hspace*{-0.8pt}D(W)$, which forms part of Kitaev's phase estimation scheme~\cite{Kitaev1996}. Registers~$\mathrm{I}$ and~$\mathrm{I\hspace*{-1pt}I}$ are complemented by~$(n+1)$ ancilla qubits, here labeled~Aux\,$0$,\,Aux\,$1$,\,\ldots,\,Aux\,$n$\,, which are all initialized in the state~$\ket{0}_{\!\mathrm{A\hspace*{-0.5pt}ux}\,i}$\ $(i=0,\ldots,n)$, followed by Hadamard gates~$H_{\!\mathrm{A\hspace*{-0.5pt}ux}\,i}$. The executions of the $(2^{m})$-th power of~$W_{k}$ is then conditioned on the state of qubit~Aux\,$m$\,, before another Hadamard gate is performed. In Fig.~\ref{fig:phase detection circuit}~(b) the approximate reflection operator (ARO) is combined from the phase detection circuit~$P\hspace*{-0.8pt}D(W)$ and its inverse~$P\hspace*{-0.8pt}D(W)^{\dagger}$, with an intermediate reflection over the ancilla state~$\ket{00\ldots0}_{\mathrm{A\hspace*{-0.5pt}ux}}$.}
\end{figure}

The next step in the design of a quantum RPS agent is the construction of the \emph{approximate reflection operator} (ARO) from the walk operator~$W(P)$. The ARO operator is designed to approximate the (ideal) reflection operator
\begin{align}
    \operatorname{ref}(\ket{\piP{P}\pr})    &=\,2\,\ket{\piP{P}\pr}\!\!\bra{\piP{P}\pr}\,-\,\mathds{1}_{\!\mathrm{I},\mathrm{I\hspace*{-1pt}I}}\,.
    \label{eq:ideal reflection operator}
\end{align}
With the generalized walk operator~$W(P)$ at hand, an approximate reflection over~$\ket{\piP{P}}$ is obtained~\cite{MagniezNayakRolandSantha2011} by implementing the phase detection operator~$P\hspace*{-0.8pt}D(W)$,~a modification of Kitaev's~\cite{Kitaev1996} phase estimation algorithm, shown in Fig.~\ref{fig:phase detection circuit}. For this task, we add~$(n+1)$ ancilla qubits, where~$n$ scales as $\log_{2}(1/\sqrt{\delta})$, where~$\delta=1-|\lambda_{2}|$ is the spectral gap of the Markov chain, i.e., $\lambda_{2}$ is the second largest eigenvalue of~$P$. We employ~$P\hspace*{-0.8pt}D(W)$ and its inverse operation, with an intermediate reflection over the ancilla state~$\ket{00\ldots0}_{\mathrm{A\hspace*{-0.5pt}ux}}\,$. This combination of operations approximates the reflection over~$\ket{\piP{P}\pr}$ from Eq.~(\ref{eq:ideal reflection operator}). An analysis of the fidelity of the reflection, as a function of~$n$, is given in Ref.~\cite{MagniezNayakRolandSantha2011}. The crucial feature of this construction is that the ARO operates based on~a number of calls to~$W(P)$ that scales as~$\tilde{O}(1/\sqrt{\delta})$ \footnote{The tilde-O notation designates that, for this analysis, we are ignoring factors which are contributing only logarithmically.}, while the number of calls to~$P$ to prepare the stationary distribution for the classical RPS scales as~$\tilde{O}(1/\delta)$.

\subsection{Quantum deliberation \label{subsec:the quantum speed-up}}

To output~a distribution of actions that corresponds to the tail of the stationary distribution with support only over the (flagged) actions, the agent performs~a quantum deliberation process with elements reminiscent of Grover-like steps~\cite{Grover1996,MagniezNayakRolandSantha2011}. In the preparation phase, the agent first initializes the joint system of registers~$\mathrm{I}$ and~$\mathrm{I\hspace*{-1pt}I}$ in the state~$\ket{\piP{P}\pr}$ from Eq.~(\ref{eq:stationary state pi pr}). While the preparation of this initial state may be involved in general, in certain cases, including the one presented in the appendix, it becomes straightforward. Consecutively, the agent alternatingly applies the following two operations:
\begin{enumerate}[\hspace*{2mm}(i)]
\item{\label{item:Grover I}Reflection over the actions:
    \begin{align}
        \operatorname{ref}(\mathcal{A})   &=\,2\sum\limits_{i\in \mathcal{A}}\ket{c_{i}}\!\!\bra{c_{i}}_{\mathrm{I}}\,-\,\mathds{1}_{\raisebox{-1.5pt}{\scriptsize{$\mathrm{I}$}}}\,,
        \label{eq:reflection over flagged}
    \end{align}
    where~$\mathcal{A}$ denotes the set of (flagged) actions.}
\item{\label{item:Grover II} Approximate reflection over the state~$\ket{\piP{P}\pr}$.}
\end{enumerate}
The sequence of operations above will, similarly to Grover's algorithm, increase the amplitude of the actions with respect to non-action components in the state of the system, while maintaining the relative weights of the action elements. This ensures that the actions are output according to the correct distribution, as explained in \cite{PaparoDunjkoMakmalMartinDelgadoBriegel2014}.

After iterating these steps~a number of times that is determined by the relative probability $\epsilon=\sum_{i\in \mathcal{A}}(\piP{P})_{i}$ of the actions within the stationary distribution, the agent samples, that is, measures in the clip basis of register~$\mathrm{I}$. If~a desired action is found, it is coupled out, otherwise the procedure is repeated~\cite{PaparoDunjkoMakmalMartinDelgadoBriegel2014}. The average number of iterations of the Grover-like steps~(\ref{item:Grover I}) and~(\ref{item:Grover II}) scales as~$\tilde{O}(1/\sqrt{\epsilon})$, while the classical RPS agent requires~$\tilde{O}(1/\epsilon)$ iterations on average.

\subsection{Reflecting PS Implementation for Trapped Ions}\label{subsec:trapped ion speed up implementation}
Finally, let us examine the possibility to implement the decision-making process of~a quantum RPS agent in an ion trap. As we have explained, two operators are required, the reflection over (flagged) actions, and the ARO. The former can be generically achieved, for instance, by applying the detuned pulses corresponding to~$U_{\!Y_{i}}^{\raisebox{-1pt}{\tiny{$S_{g}$}}}(2\pi)$ or~$U_{\!Y_{i}}^{\raisebox{-1pt}{\tiny{$S_{g}$}}}(2\pi)$ of the coherent controlization step~(\ref{item:adding control iv}) specifically to those basis states corresponding to (flagged) actions, flipping their sign. The latter, the ARO, is implemented starting from the probability unitaries, by coherent controlization, in conjunction with~a few fixed operations, $D_{0,\mathrm{I}}$, $D_{0,\mathrm{I\hspace*{-0.5pt}I}}$,~$D_{0,\mathrm{A\hspace*{-0.5pt}ux}}$ and~$H$.

Let us briefly describe the individual steps of this procedure. By coherently conditioning the probability unitaries~$U_{i}$, the operation~$U_{\!P}$ is obtained, from which the pulse sequence for~$V_{\!P}$ is obtained by swapping the registers, which, in practice, corresponds to an exchange of the qubit/ion labels in the pulse sequence for~$U_{\!P}$. The associated inverse operators follow immediately by setting~$\theta_{i}\rightarrow(-\theta_{i})\,$. The reflections $D_{0,\mathrm{I}}$, $D_{0,\mathrm{I\hspace*{-0.5pt}I}}$, and~$D_{0,\mathrm{A\hspace*{-0.5pt}ux}}$ are obtained as special cases of the reflection over the (flagged) actions. The Hadamard gate
\begin{align}
    H   &=\,\frac{1}{\sqrt{2}}\begin{pmatrix}
        1   &  \ 1\\
        1   &   -1
    \end{pmatrix}\,,
    \label{eq:Hadamard}
\end{align}
can be implemented up to~a phase of~$(-i)$, that is, for the $j$-th ion we have the pulse sequence
\begin{align}
    -i\,H   &=\,U_{\!X}(-\tfrac{\pi}{2})\,U_{Z_{j}}(\tfrac{\pi}{2})\,U_{\!X}(\tfrac{\pi}{2})\,U_{Z_{j}}(\pi)\,,
    \label{eq:Hadamard pulse decomposition}
\end{align}
with~$U_{\!X}$ as in Eq.~(\ref{eq:collective X rotation gate}), and~$U_{Z_{j}}$ given by Eq.~(\ref{eq:single qubit Z gate}). The superfluous phase~$(-i)$ cancels naturally, since the Hadamard gate is used four times for every ancilla in the ARO, twice each for the realization of~$P\hspace*{-0.8pt}D(W)$ and its inverse, see Fig.~\ref{fig:phase detection circuit}. Finally, we make again use of coherent controlization to construct the phase detection operator~$P\hspace*{-0.8pt}D(W)$ and its inverse from the walk operator~$W(P)$. The possibility to add control to arbitrary (unknown) unitaries hence provides~a modular structure, that allows, in principle, for the generic implementation of all operations that required for the decision-making of~a quantum RPS agent. The modular use of coherent controlization in the design of the agent can thus be summarized by the following sequence:
\begin{align}
    U_{Y}(\theta_{i})   \stackrel{CC}{\longrightarrow}
    U_{j}(\theta_{1},\ldots,\theta_{N-1})   \stackrel{CC}{\longrightarrow}
    U_{\!P},W(P)
    \stackrel{CC}{\longrightarrow}
    \mathrm{ARO}.
    \nonumber
\end{align}
That is, starting from single qubit $Y$~rotations, parameterized according to the stochastic matrix~$P$, we construct the probability unitaries using coherent controlization. From the probability unitaries we then construct, again by coherent controlization, $U_{\!P}$ and $V_{\!P}\,$, which are used to assemble~$W(P)$. Finally, from~$W(P)$ we construct the ARO operator that is central to the quantum deliberation steps, once again employing coherent controlization.

As we have argued, all individual operations of the quantum RPS are implementable with current technology. While large network sizes, as well as small values of~$\epsilon$ or~$\delta$, impose challenges for state-of-the-art ionic implementations of the generic RPS decision-making process, these technological restrictions may be overcome by the continuing development of scalable ion trap arrays. Nonetheless, special cases of the general scheme we have laid out here are well within reach of experimental testing. In the \hyperref[Appendix]{Appendix}, we present such an example for~a quantum RPS agent based on an ECM using two qubits, and we give an explicit pulse decomposition of its entire decision-making process, including an error analysis.\\
\vspace*{-8mm}

\section{Conclusions}\label{sec:conclusions}
We have presented a modular architecture for the implementation of the deliberation process of PS agents in systems of trapped ions. We have shown first how the probability unitaries, which are required for the emulation of classical random walks, can be generically constructed using coherent controlization, and second how this process allows for the implementation of~a quantum RPS agent based on these probability unitaries. A~main feature of our construction is its modular architecture, that is, any changes of the probabilities as part of the learning process can be dealt with at the level of the implementation of the probability unitaries, whereas the rest of the construction is unaltered. The generic construction relies only on elementary single-qubit~$Y$ rotations and coherent controlization, which allows for~a straightforward assembly, as well as straightforward updating of the probability unitaries.

This is an important advantage, if not~a prerequisite, for the realization of~a learning agent that is continuously adjusting the probabilities underlying its deliberation process. Having to re-compute the entire sequence of gates which need to be applied to realize the quantum RPS agent for any change of the underlying Markov chain would impose~a large computational overhead on the agent, and significantly diminish the advantage in speed that is provided by quantizing the RPS agent.

In addition to the general modular architecture, we have provided numerical simulations of an implementation of simple RPS agents using trapped ions. As our investigation shows, proof-of-principle realizations of these agents are simple enough to be implementable in current experimental setups, while they are sufficiently involved to demonstrate the quadratic speed-up.


\begin{acknowledgements}
\vspace*{-2mm}
We are grateful to Adi Makmal, Markus Tiersch, Benjamin~P. Lanyon, Daniel Nigg and Thomas Monz for valuable discussions and comments. HJB acknowledges discussions with Gavin Brennen at an early stage of this project. This work has been supported in part by the Austrian Science Fund (FWF) through the SFB FoQuS: F4012 and the Templeton World Charity fund grant TWCF0078/AB46.
\end{acknowledgements}

\appendix*

\section{Rank-One Reflecting PS in Ion Traps}\label{Appendix}
Here, we provide an example for~a quantum RPS agent sophisticated enough for the demonstration of~a quantum speed-up, whilst being sufficiently simple to allow an immediate implementation in readily available ion trap setups, e.g., as described in Refs.~\cite{Schindler-Blatt2013,Barreiro-Blatt2011}. The Appendix is structured as follows. In Section~\hyperref[subsec:rank one RPS]{1} we first discuss the simplified decision-making process for~a quantum RPS agent whose underlying ECM network corresponds to a rank-one Markov chain. To provide context, the role of these simple agents is then illustrated for the invasion game in Section~ \ref{subsec:invasion game}. In Section~\ref{subsec:Rank-One Quantum RPS with Trapped Ions}, we propose an ion trap implementation of the rank-one quantum RPS agent, for which we supply the explicit overall pulse sequence. We accompany our proposal with an appropriate error model, and corresponding numerical simulations, which are given in the final Section~\ref{subsec:Numerical simulations}.\\
 \vspace*{-6mm}

\subsection{Rank-One Reflecting PS}\label{subsec:rank one RPS}

A~special case of the RPS agents that we have considered in Section~\ref{sec:reflecting projective simulators in ion traps} is obtained by considering the reflective analog of so-called ``two-layered" PS agents, where all transition are one-step transitions from percepts to actions~\cite{PaparoDunjkoMakmalMartinDelgadoBriegel2014}. Such agents have a very simple structure, yet were shown to be capable of learning to solve non-trivial environmental tasks~\cite{MautnerMakmalManzanoTierschBriegel2014,MelnikovMakmalBriegel2014}. In the RPS analog of two-layered PS agents \cite{PaparoDunjkoMakmalMartinDelgadoBriegel2014}, the associated Markov chains of each percept-specific clip network are rank-one throughout the entire learning process of the agent. The columns of~$P$ are then all identical, and equal to the stationary distribution. The spectral gap is given by~$\delta=1$, and the Markov chain mixes in one step. Let us consider the consequences\textemdash radical simplifications\textemdash for the construction of the RPS agent.

\begin{figure}[ht!]
(a)\includegraphics[width=0.42\textwidth]{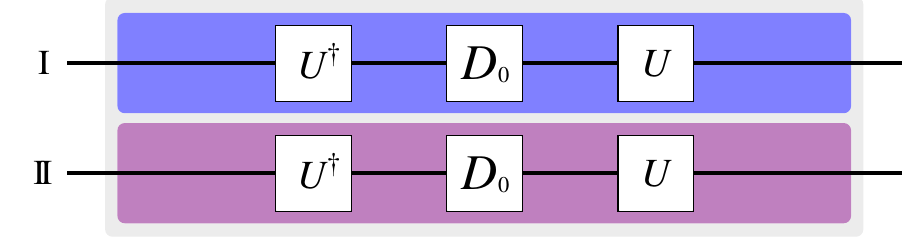}
(b)\includegraphics[width=0.42\textwidth]{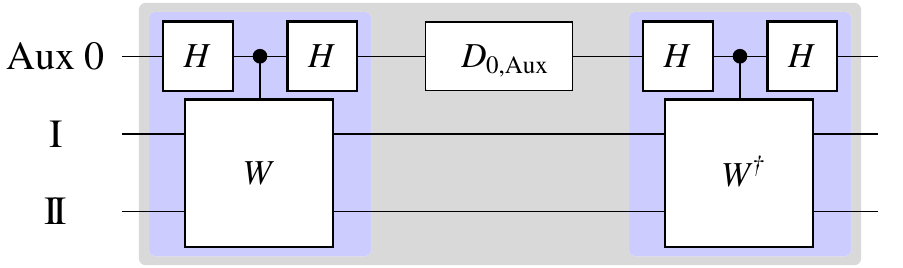}
(c)\includegraphics[width=0.42\textwidth]{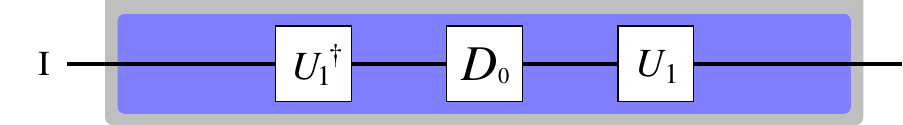}
\caption{\label{fig:rank one walk operator}
\textbf{Rank-one reflection operator}. For rank-one Markov chains, $U_{\!P}$ and~$V_{\!P}$ are local operations on registers~$\mathrm{I\hspace*{-0.5pt}I}$ and~$\mathrm{I}$, respectively. The Szegedy walk operator~$W(P)$ that is shown in Fig.~\ref{fig:rank one walk operator}~(a) hence factorizes into two independent applications of~$U\,D_{0}\,U^{\dagger}$. Since the walk operator further becomes Hermitean, $W=W^{\dagger}$, the single remaining ancilla is also redundant, the approximate reflection circuit shown in Fig.~\ref{fig:rank one walk operator}~(b) reduces to one application of~$W(P)$ as shown in Fig.~\ref{fig:rank one walk operator}~(c), and the reflection becomes exact.}
\end{figure}

In the rank-one case, the probability unitaries~$U_{i}$ for~a fixed~$P$ are all the same, so we may remove the subscript, write only~$U$, but we keep in mind the distinction of~$U$ and~$U_{\!P}\,$. Moreover, coherent controlization is no longer necessary for the construction of~$U_{\!P}$, since~$U$ is applied regardless of the state of the control register, $U_{\!P}=\mathds{1}\otimes U$ ($V_{\!P}=U\otimes\mathds{1}$). As can be easily seen, the reflections~$\operatorname{ref}(A)$ and~$\operatorname{ref}(B)$ shown in Fig.~\ref{fig:generalized walk operator} then commute, acting locally on registers~$\mathrm{I\hspace*{-0.5pt}I}$ and~$\mathrm{I}$, respectively, see Fig.~\ref{fig:rank one walk operator}. Similarly, the coherent encoding of the stationary distribution is now given by the product state~$\ket{\piP{P}\pr}_{\mathrm{I},\mathrm{I\hspace*{-1pt}I}}=\ket{\piP{P}}_{\mathrm{I}}\ket{\piP{P}}_{\mathrm{I\hspace*{-1pt}I}}\,$.

When assembling the phase detection operator~$P\hspace*{-0.8pt}D(W)$ and the approximate reflection operator (ARO), see Fig.~\ref{fig:phase detection circuit}, the spectral gap of~$\delta=1$ means that (at most) one ancilla qubit is required. Now, note that the walk operator~$W(P)$ for rank-one matrices~$P$, as shown in Fig.~\ref{fig:rank one walk operator}~(a), is Hermitean, and thus the entire circuit shown in Fig.~\ref{fig:rank one walk operator}~(b) reduces to~a single application of the Szegedy walk operator~$W(P)$. An \emph{exact} reflection over~$\ket{\piP{P}}$ can hence be performed by applying $W(P)=U\,D_{0}\,U^{\dagger}$ to either of the registers, see Fig.~\ref{fig:rank one walk operator}~(c). Without loss of generality we select register~$\mathrm{I}$, where we drop the subscript indicating the register from now on, to perform all the Grover-like steps to output actions according to the tailed stationary distribution, which entails the following steps.

In the preparation stage, the state~$\ket{\piP{P}}$ is initialized by one application of~$U$ to the state~$\ket{0}$. Then, the two operators of the Grover-like process, i.e., the reflection over the action $\operatorname{ref}(\mathcal{A})$, and the reflection over~$\ket{\piP{P}}$, are applied~a prescribed number of times determined by~$\epsilon$, the relative probability of the actions within the stationary distribution. Consecutively, the agent measures in the clip basis. If the measurement provides an action, it is coupled out, otherwise the agent iterates this procedure.

Before we continue with the ionic implementation of the deliberation process, let us briefly examine an example for~a task\textemdash the invasion game\textemdash for which the agent may employ its capabilities of learning and decision-making.

\subsection{The Invasion Game}\label{subsec:invasion game}

\begin{figure}[ht!]
\includegraphics[width=0.49\textwidth]{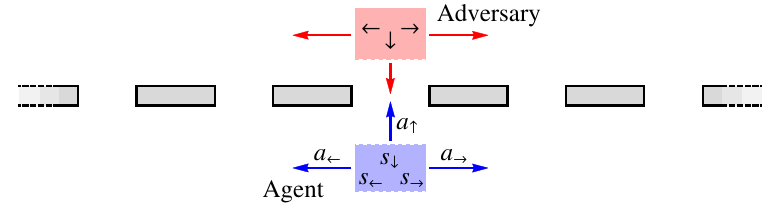}
\caption{\label{fig:invasion game}
\textbf{Invasion game}. In the invasion game~\cite{BriegelDeLasCuevas2012} the agent defends~a region of space against an adversary that tries to enter through~a series of openings. To be rewarded, the agent is to prevent the adversary from entering, by blocking the passages, which can be achieved if the adversary's signals,~``$\downarrow$", ``$\leftarrow$", and~``$\rightarrow$", indicating its next move, are interpreted correctly, and the agent mirrors the adversary's moves.}
\end{figure}

As~a simple example that can be solved by two-layered agents, let us discuss the invasion game, as considered in Ref.~\cite{BriegelDeLasCuevas2012}. In this game, the agent is tasked with guarding~a region of space from an adversary who attempts to enter the region through an array of entrances, see Fig.~\ref{fig:invasion game}. The agent's goal is to prevent the adversary from entering by blocking sites. In every round of the game, the adversary has three possible moves. It may attempt to enter at its current location, or move one door to the left, or one door to the right and attempt to enter through one of these openings. The agent is rewarded if it matches the move, thus blocking the adversary.

To emphasize the learning aspect of the game, we assume that the game starts with the adversary and the agent located at the same entrance, and before the adversary moves, it displays some signal that indicates which way he intends to move next. Thus, the set of percepts of the agent (the defender) is $\{\downarrow, \leftarrow, \rightarrow\}$, which hint at the possible subsequent move of the attacker. The agent itself can also choose to remain where it is, move left, or move right in an attempt to block, corresponding to the three action clips $c_{1}(a_{\downarrow})\,$,~$c_{2}(a_{\leftarrow})\,$, and~$c_{3}(a_{\rightarrow})\,$ accessible to the agent.

For the RPS agents discussed previously, this simple game may be represented by associating~a three clip network to each of the percepts.
In what follows, we shall only focus on~a network associated to one percept, say ``$\downarrow$", as everything will also hold for other subnetworks as well, and we shall drop the corresponding subscript for ease of notation. For such two-layered settings there is~a simple construction relating the probabilities of outputting~a particular action, and the structure of the underlying percept-specific Markov chain. In particular, the action probabilities $\pi = (\pi_{1}, \pi_{2}, \pi_{3})$ are realized by the stochastic matrix where  each column is the vector~$\pi$. The learning of the agent manifests in the relative increases of probabilities corresponding to rewarded actions, and examples for specific update rules can be found, e.g., in Ref.~\cite{BriegelDeLasCuevas2012}.

In basic two-layered settings in both the RPS and the analogous standard PS agent models, an action is coupled out after exactly one diffusion step. In order to illustrate~a speed-up in such~a scenario, we therefore need to consider some additional structure that increases the learning efficiency of the agent, but induces~a longer deliberation time. Such~a structure can be provided by percept-specific \emph{flags}, which correspond to rudimentary emotion tags. Flags can be interpreted as the agent's short term memory, indicating favored actions. In other words, absent flags indicate that~a particular choice of action, for~a given percept, was not rewarded in the previous step, and should be avoided. More precisely, this structure works as follows. Initially, all the actions are flagged. Then after an action has been coupled out, the flag is removed if the action is not rewarded. If the unflagged action is selected again after encountering the
\vspace*{-2mm}
same percept in~a consecutive round, the deliberation process is repeated until the deliberation results in~a flagged action. In the case that the last remaining flag is removed, which indicates~a definite change in the setting of the environment, all flags are re-set.

This structure leads to great improvements in settings where the environment (e.g., the adversary in the invasion game) changes its strategy, for instance, by permuting the meaning of the percepts \cite{BriegelDeLasCuevas2012}. In this case, if the network is already well-taught, the probability of outputting the correct action, once the meaning of percepts has been altered,  can be very low. We will be interested in precisely such~a setting. Suppose the attacker pursues~a consistent strategy for~a prolonged period of time, and the agent has learned well. This entails that, for a given percept, one of the values in the distribution $\pi = (\pi_{1}, \pi_{2}, \pi_{3})$, say the third, is much larger than the others, e.g., $\pi_{3}\ll \epsilon=\pi_{1}+\pi_{2}\,$, and only the action clip corresponding to~$\pi_{3}$ is flagged. Now, if the environment is to suddenly change its strategy, no longer rewarding this action, the flag on this clip will disappear, while flags on other clips are introduced again. Subsequently, the agent is required to output the tail of the distribution~$\pi$ with support only over the actions corresponding to~$\pi_{1}$ and $\pi_{2}$. However, for the classical RPS model, as well as for the standard PS model, the average number of iterated diffusion steps required until one of the remaining flagged actions is hit is $O(1/\epsilon)$, which can be exceptionally large, if the network was well-taught. The quantum variant of the RPS will then be quadratically faster, only requiring $O(1/\sqrt{\epsilon})$ steps. In any given round, the decision-making process after encountering~a percept can then be represented on~a two-qubit Hilbert space according to Table~\ref{table:four clip network}.

\begin{table}[h!]
\setlength{\tabcolsep}{7pt}
\renewcommand{\arraystretch}{1.5}
\begin{tabular}{c c c}
    \hline\hline
    clip    &   interpretation  &   two-qubit state\\   \hline
    $c_{1}$   &   action\ $a_{\downarrow}$&   $\ket{00}$ \\  \hline
    $c_{2}$   &   action\ $a_{\leftarrow}$&   $\ket{01}$ \\  \hline
    $c_{3}$   &   action\ $a_{\rightarrow}$ &   $\ket{10}$, $\ket{11}$ \\  
    \hline\hline
\end{tabular}
\caption{Representation of three-clip network as two qubits. A two qubit system can represent four clips, but as the desired network only requires three, a redundancy is introduced, e.g. in clip $c_3$.}
\label{table:four clip network}
\end{table}

Next, we discuss how~a rank-one quantum RPS deliberation process based on this two-qubit system can be represented using two trapped ions.

\subsection{Rank-One Quantum RPS with Trapped Ions}\label{subsec:Rank-One Quantum RPS with Trapped Ions}

To implement~a rank-one quantum RPS agent for~a setting such as the one described above, we construct the two-qubit operations~$U$, $D_{0}$ and~$\operatorname{ref}(\mathcal{A})$, where the latter operation is now~a reflection over flagged actions only, from laser pulses on two trapped ions. As we have described in Section~\ref{subsec:Coherent Controlization}, coherent controlization may be employed to assemble the probability unitary~$U$, but in this simple case we may resort to~a simpler option. As shown in Table~\ref{table:four clip network}, we operate on~a two-qubit Hilbert space, but we only distinguish between three clips, such that only two independent angles,~$\theta_{1}$ and~$\theta_{2}\,$, parameterize the probability unitary~$U$. A~pulse sequence that achieves this is given by

\begin{align}
    U(\theta_{1},\theta_{2})    &=\,
    U_{\!X}(-\tfrac{\pi}{2})\,U_{\!Z_{2}}(2\theta_{2})\,U_{\!Z_{1}}(2\theta_{1})\,U_{\!X}(\tfrac{\pi}{2})\,,
    \label{eq:three clip prob unitary pulses}
\end{align}
where the collective~$X$ and single-qubit~$Z$ pulses are realized by individual laser pulses as described in Section~\ref{subsec:standard PS with trapped ions}. In terms of the probabilities~$\pi_{1}$ and~$\pi_{2}$, which we assume correspond to the two flagged actions, the angles $\theta_{1}$ and~$\theta_{2}$ are given by
\begin{subequations}
\label{eq:angles vs probabilities in three clip network}
\begin{align}
    \theta_{1}  &=\,\arccos\sqrt{\pi_{1}+\pi_{2}}\,,
    \label{eq:theta 1 in terms of pi1 and pi2}\\
    \theta_{2}  &=\,
    \begin{cases}\displaystyle
    \arccos\sqrt{\frac{\pi_{1}}{\pi_{1}+\pi_{2}}}\,,   &   \mbox{for}\ \pi_{1}+\pi_{2}\neq0\,,\\
    0\,,   &   \mbox{otherwise}\,.
    \end{cases}
\end{align}
\end{subequations}

For the implementation of~$\operatorname{ref}(\mathcal{A})$, the reflection over the actions, one simply applies the single-qubit~$Z$ operation \begin{align}
    U_{\!Z_{1}}(\pi)    &=\,\diag\{-i,-i,i,i\}\,.
    \label{eq:simple reflection over flagged}
\end{align}
Since the rank-one RPS operates solely on one register, the overall phase of the reflection is irrelevant, as long as the relative sign between flagged actions and all other clips is flipped. Finally, we propose the following implementation of~$D_{0}\,$. A~detuned M{\o}lmer-S{\o}rensen pulse, see Ref.~\cite{MoelmerSoerensen1999}, is used to transfer the population of the state~$\ket{gg}$, corresponding to~$\ket{00}$, to an auxiliary state~$\ket{g\pr g\pr}$. While the state~$\ket{00}$ is hidden in this way,~a single-qubit~$Z$ pulse~$U_{\!Z_{1}}(2\pi)$ flips the sign of all other basis states, before~a second M{\o}lmer-S{\o}rensen pulse returns the population to~$\ket{00}$.

Taken together, all operations for one iteration of the Grover-like reflection may hence be realized by~$12$ laser pulses. In addition,~$4$ individual pulses are needed for the preparation of the initial state. At last, in the next section, we investigate the performance of our ion-trap quantum RPS agent in~a series of numerical simulations that incorporate~a suitable error model.

\vspace*{-2mm}
\subsection{Numerical simulations}\label{subsec:Numerical simulations}

For the numerical simulations that we present in this final section, 
we consider imprecisions in the laser pulse frequency or duration, resulting in varying angles for the laser pulses, as the primary sources of errors. We model such errors by randomly varying the angles for each pulse in the sequence according to~a Gaussian distribution with standard deviation~$\sigma$ that is centered around the correct value. 

In the simulations, we specify~a pair of values~$\pi_{1}>0$ and~$\pi_{2}>0$, such that~$\epsilon=\pi_{1}+\pi_{2}<1$, initialize the corresponding state vector~$\ket{\piP{P}}=U(\theta_{1},\theta_{2})\ket{0}$, and apply the combination of the reflections~$\operatorname{ref}(\mathcal{A})$ and $U\,D_{0}\,U^{\dagger}$ a~total of~$m$ times, where~$m\in\mathbb{N}$ is chosen randomly from the interval~$\left[0,m_{\epsilon}\right]$, with ~$m_{\epsilon} = \lceil 1/\sqrt{\epsilon} \rceil$. The clips are then randomly sampled according to the probability distribution
\begin{align}
    \left\{\bigl|\hspace*{-0.5pt}\bra{c_{i}}
    \bigl[U\,D_{0}\,U^{\dagger}\operatorname{ref}(\mathcal{A})\bigr]^{m}\,U\ket{0}\hspace*{-0.5pt}\bigr|^{2}
    \right\}_{i}\,,
    \label{eq:prob distr simulations}
\end{align}
which corresponds to~a measurement in the clip basis. If no flagged action is found,~a new number~$m$ is generated, and the procedure is iterated until~a flagged action has been sampled. For every fixed set of~$\pi_{1}$ and~$\pi_{2}$ the process is repeated for~$10^{4}$ runs to build up statistics, out of which~$N_{1}$ ($N_{2}$) result in an output of the action clip~$c_{1}$ ($c_{2}$), corresponding to~$\pi_{1}$ ($\pi_{2}$). Additionally, the overall number~$N_{U}$ of calls to the operator~$U$ until~a flagged action is observed is recorded in each run.

\begin{figure}[ht!]
(a)\includegraphics[width=0.43\textwidth]{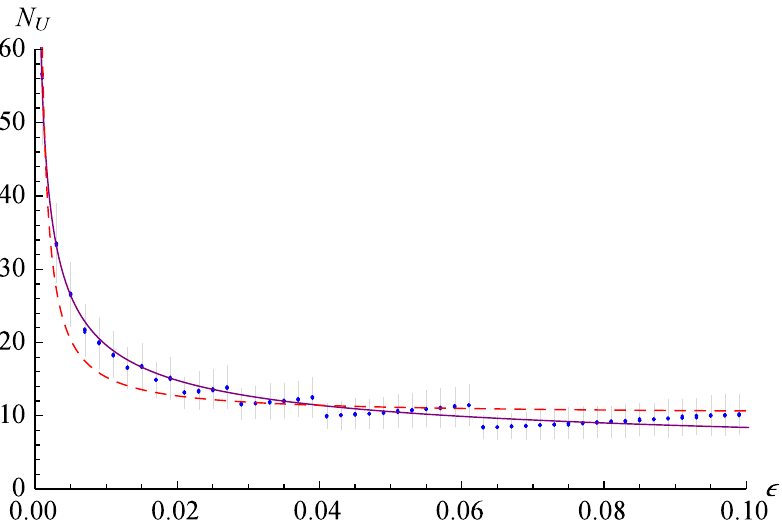}
(b)\includegraphics[width=0.43\textwidth]{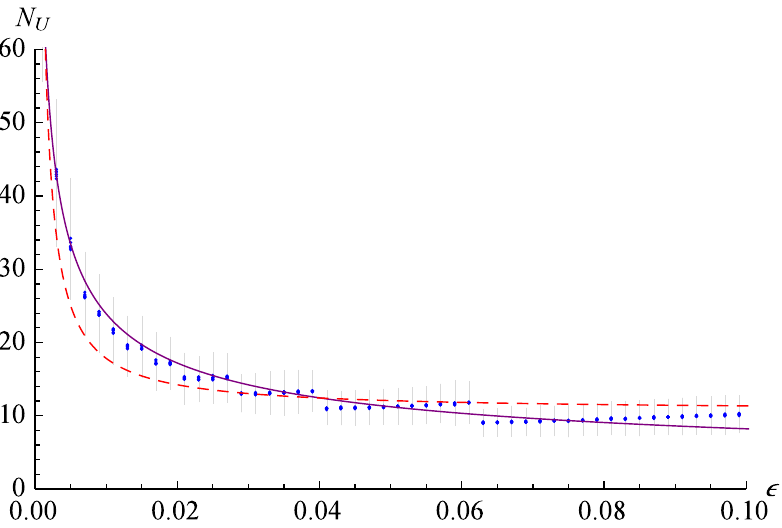}
(c)\includegraphics[width=0.43\textwidth]{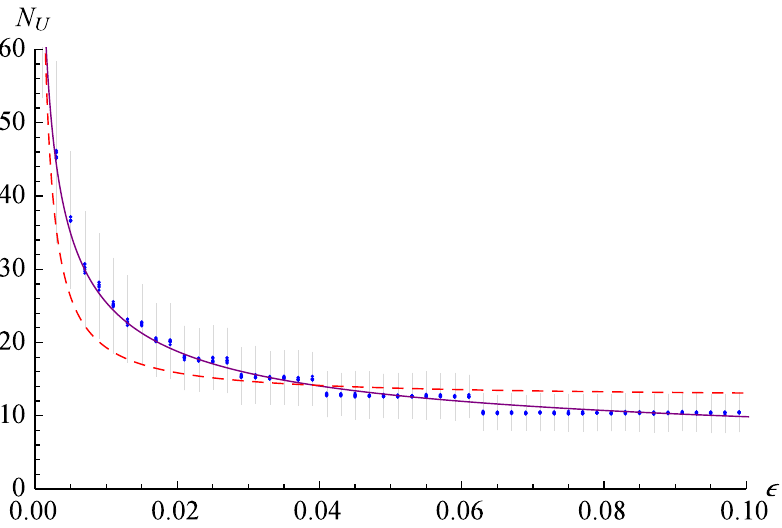}
\caption{\label{fig:NU_again_epsilon}
\textbf{Average number of calls to $U$}. The results of the numerical simulation for the average number of calls to the probability unitary~$U$ until an action clip is hit are shown for error parameters~$\sigma=\pi/100$, $\sigma=\pi/20$, and~$\sigma=\pi/10$, in Figs.~\ref{fig:NU_again_epsilon}~(a),~(b), and~(c), respectively. Each blue dot corresponds to the average over $10^{4}$ runs for~a fixed value~$\epsilon=\pi_{1}+\pi_{2}$. The vertical gray lines indicate three standard deviations of the mean values (over $100$ runs each) in each direction. The solid purple curves show the best fits that are linear in $(1/\sqrt{\epsilon})$, while the dashed red curves show the best fits that are linear in $(1/\epsilon)$, and we have confirmed that the former fit the data better than the latter.}
\end{figure}

For~$N_{U}$ the expected scaling as~$(1/\sqrt{\epsilon})$ is largely independent from the error parameter, as can be seen from Fig.~\ref{fig:NU_again_epsilon}, since this behavior is governed by the structure of the process, in particular, the upper bound~$m_{\epsilon}$ for the randomly chosen value~$m$. The integer steps by which~$m_{\epsilon}$ increases, as~$(1/\sqrt{\epsilon})$ decreases, also explain the step-like pattern visible in the data of Fig.~\ref{fig:NU_again_epsilon}. That is, in such~a Grover-like scheme, the probability to sample~a flagged action grows monotonically with the number of iterations only up to some point, from which on additional applications of the reflections will alternatingly decrease and increase the probability. The average number of repetitions set by the value~$m_{\epsilon}$, which corresponds to~a fixed interval of $\epsilon$-values, is hence not optimal for all~$\epsilon$ within that interval, which can be seen from the slanting of the data points, and their standard deviations, in each of the `steps' seen in Fig.~\ref{fig:NU_again_epsilon}~(a). The errors partially cover this effect, as can be seen in Figs.~\ref{fig:NU_again_epsilon}~(b) and~(c).

\begin{figure}[ht!]
\includegraphics[width=0.43\textwidth]{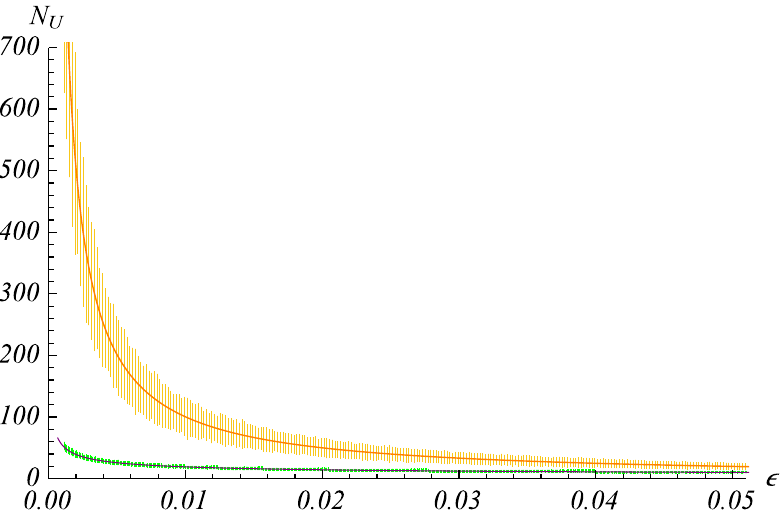}
\caption{\label{fig:classical vs quantum RPS}
\textbf{Comparison of classical and quantum RPS}. Numerical simulations of classical (upper data points) and quantum RPS (lower data points) agents are shown. The data points are obtained as averages over $10^{4}$ runs for each value of~$\epsilon$. The fitted curves that are shown are linear in $(1/\epsilon)$ (top orange curve) and $(1/\sqrt{\epsilon})$ (bottom purple curve), respectively. The vertical green and yellow bars indicate three standard deviations of the mean values (over $100$ runs each) in each direction. }
\end{figure}

To illustrate the speed-up of the quantum RPS agent with respect to~a classical RPS agent, we directly compare their performance in~a simulation without errors, that is, for~$\sigma=0$, see Fig.~\ref{fig:classical vs quantum RPS}. The classical rank-one RPS agent is emulated here by running the rank-one quantum RPS deliberation process described in this section for~$m_{\epsilon}=0\,$, that is, the state $U\ket{0}$ is prepared, and~a sample is taken, such that clip~$c_{i}$ is obtained with probability $|\bra{c_{i}U\ket0}|^{2}$. If no flagged action is obtained, the procedure is repeated.

\begin{figure}[ht!]
\includegraphics[width=0.49\textwidth]{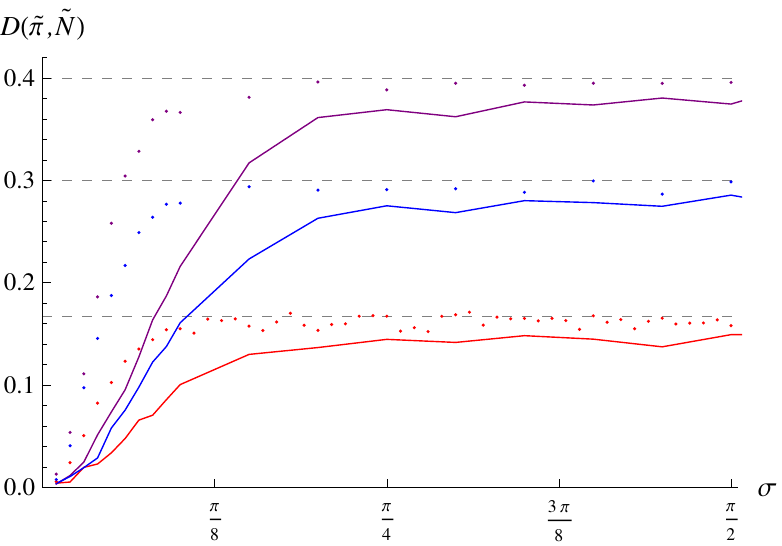}
\caption{\label{fig:statistical distance against sigma}
\textbf{Statistical distance to tailed distribution}. The statistical distance~$D(\tilde{\pi},\tilde{N})$, see Eq.~(\ref{eq:statistical distance from stationary}), of the output from the tailed stationary distribution is plotted against the width~$\sigma$ of the error distribution, for values~$\epsilon=0.05$ (solid) and $0.001$ (dots), and ratios~$\pi_{1}/\pi_{2}=9$,~$4$, and~$2$ (top to bottom). The dashed horizontal lines indicate the statistical distance to the uniform distribution for each pair~$\{\pi_{1},\pi_{2}\}$, which is approached when the errors dominate the behavior of the agent.}
\end{figure}

\begin{figure}
(a)\includegraphics[width=0.49\textwidth]{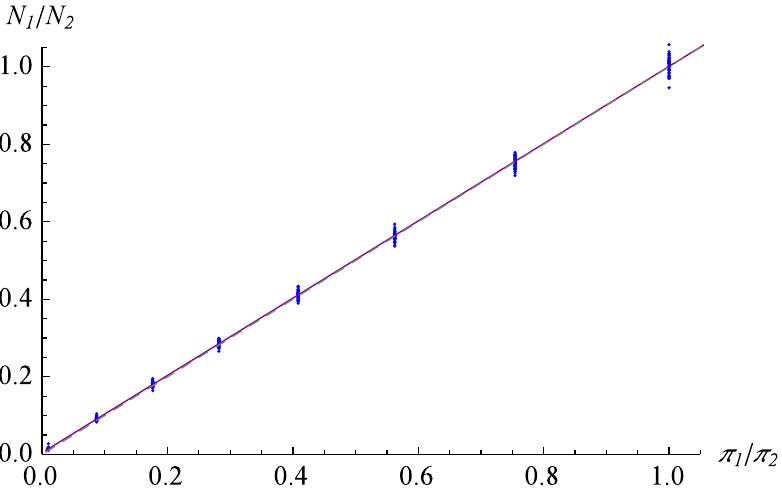}
(b)\includegraphics[width=0.49\textwidth]{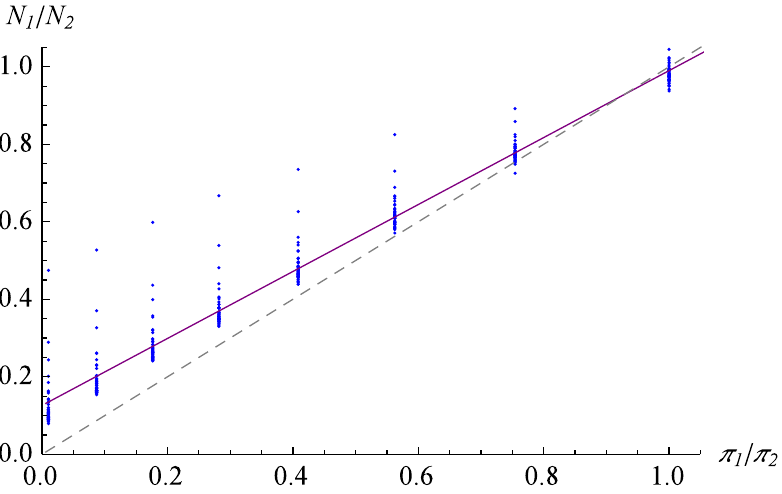}
(c)\includegraphics[width=0.49\textwidth]{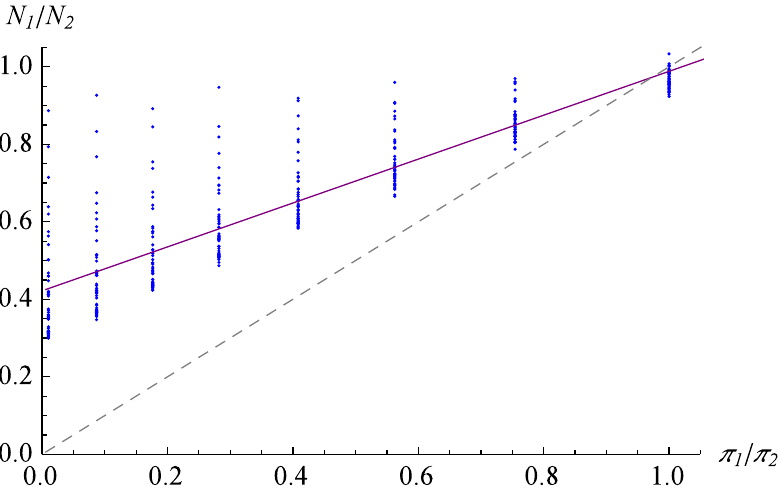}
\caption{\label{fig:output according to stationary}
\textbf{Output according to tailed distribution}. The plots in Fig.~\ref{fig:output according to stationary}~(a),~(b), and~(c) show the ratios~$N_{1}/N_{2}$ of the counts in the numerical simulations in comparison with the corresponding ratios~$\pi_{1}/\pi_{2}$ according to the (tailed) stationary distribution, for error parameters~$\sigma=\pi/100$, $\sigma=\pi/20$, and~$\sigma=\pi/10$, respectively. The solid purple lines show the best linear fits, which should match the $45°$ diagonal, shown as dashed gray line, in an ideal RPS agent. Each group of data points along~a vertical line corresponds to fixed value of~$\pi_{1}/\pi_{2}$, but varying~$\epsilon$. The data used is in fact the same as that used for Fig.~\ref{fig:NU_again_epsilon}.}
\end{figure}

What remains to be confirmed by the simulations is the output of flagged actions according to the tail of the stationary distribution, as predicted in Ref.~\cite{PaparoDunjkoMakmalMartinDelgadoBriegel2014}. We address this question in two ways. First, we evaluate the behavior of~a few selected illustrative pairs of probabilities~$\pi_{1}$ and~$\pi_{2}$ for increasing error parameters in Fig.~\ref{fig:statistical distance against sigma}. As~a measure for the accuracy of the output, we use the statistical distance
\begin{align}
    D(\tilde{\pi},\tilde{N})  &=\,\frac{1}{2}\sum_{i=1,2}\bigl|\frac{\pi_{i}}{\pi_{1}+\pi_{2}}-\frac{N_{i}}{N_{1}+N_{2}}\bigr|\,,
    \label{eq:statistical distance from stationary}
\end{align}
of the output distribution~$\tilde{N}=\{N_{i}/(N_{1}+N_{2})\}_{i=1,2}$ and the tailed stationary distribution~$\tilde{\pi}=\{\pi_{i}/(\pi_{1}+\pi_{2})\}_{i=1,2}$. In Fig.~\ref{fig:output according to stationary} we then compare the relative frequencies $N_{1}/N_{2}$ with which the two flagged actions were obtained to the corresponding ratios $\pi_{1}/\pi_{2}$ of the (tailed) stationary distribution, for~a broad range of values~$\pi_{1}$ and~$\pi_{2}$, and for the three error parameters previously chosen used in Fig.~\ref{fig:NU_again_epsilon}.

The data shown in Fig.~\ref{fig:statistical distance against sigma} illustrates that large errors result in an output according to~a uniform distribution over the flagged actions. The farther the tailed stationary distribution is away from the uniform distribution, the smaller the tolerance for errors. As the stationary distribution is updated throughout the learning process the errors will thus cause~a stronger deviation from the desired output distribution.

To make these statements more meaningful in terms of learning agents, let us consider~a specific example. Let us assume that for~a fixed percept, the tailed stationary distribution may be biased towards the action clip~$c_{1}$, such that an ideal agent outputs this action in~$90\%$ of the cases\footnote{The so-called `forgetfulness parameter' $\gamma$ which controls at what rate the agent is, roughly speaking, forgetting what it has learned, which radically speeds up re-learning \cite{MautnerMakmalManzanoTierschBriegel2014} also implies that the output efficiency is bounded below 1, and depends on $\gamma$. For our examples we opt to consider the case where this efficiency is at 0.9.}.
To reach this goal, such an agent updates the corresponding Markov chain throughout the learning process, until the associated stationary distribution is such that~$\pi_{1}/\pi_{2}=9$. We may then set an error threshold, by assuming that the agent is still considered to succeed, if the action~$c_{1}$ is performed only~$70\%$ of the time, i.e.,~a statistical distance of~$20\%$. Brief inspection of the topmost solid curve in Fig.~\ref{fig:statistical distance against sigma} reveals that for~$\epsilon=0.05$ the threshold value corresponds roughly to the largest error,~$\sigma=\pi/10$, that we consider in Fig.~\ref{fig:NU_again_epsilon}. This, in turn, suggests~a maximal number of~$m_{\epsilon}=5$ coherent iterations of the reflections in the Grover-like process before~a measurement is performed, which translates to~$64$ individual laser pulses as described in Section~\ref{subsec:Rank-One Quantum RPS with Trapped Ions}.\\

The initial analysis presented in this appendix suggests that our proposal for the implementation of two-layered quantum RPS agents may be feasible, and be readily implemented in a laboratory as a proof-of-principle demonstration of learning agents enhanced by employing quantum physics.

\clearpage

\end{document}